\documentclass[12pt,reqno]{article}
\usepackage[T1]{fontenc}
\usepackage[utf8]{inputenc}
\usepackage{newpxtext,newpxmath}
\usepackage{microtype}
\usepackage[fontsize=12pt]{fontsize}

\usepackage[right=1.08in,left=1.08in,top=1.08in,bottom=1.08in]{geometry}

\usepackage{amsmath,amsthm,graphicx}
\usepackage{subcaption}
\usepackage{booktabs,multirow,threeparttable,array,adjustbox}
\usepackage{makecell}
\usepackage{setspace}
\usepackage[dvipsnames]{xcolor}
\usepackage{natbib}
\usepackage{xurl}
\usepackage{hyperref}
\hypersetup{
colorlinks=true, linkcolor=red!75!black, citecolor=blue!75!black, filecolor=magenta, urlcolor=red!75!black,
}
\usepackage[noabbrev]{cleveref}
\usepackage{sectsty}
\usepackage{bm}
\usepackage{fancyhdr}
\title{
\vspace*{-2.5cm} \hspace*{-0.5cm}
\textsc{Watching Trade from Space: Measuring Maritime Trade Using Satellite Imagery} \thanks{Earlier versions of this paper circulated under the title ``Watching Trade from Space: Nowcasting and Spatial Extrapolation of Port-Level Maritime Trade Using Satellite Imagery''. Replication package is available at \url{https://github.com/yonggeun-jung/watching_trade_public}. I am deeply grateful to Gunnar Heins and Sooji Kim for their invaluable comments and suggestions. I also thank participants at the 11th Monash-Warwick-Zurich-CEPR Text-As-Data Workshop and the Conference of the Economic Data Science Society 2026 for helpful comments. All remaining errors are my own. First version: May, 2025.}
}

\author{
\textsc{Yonggeun Jung}\protect\thanks{Department of Economics, University of Florida. Email: \href{mailto:yonggeun.jung@ufl.edu}{yonggeun.jung@ufl.edu}.}}

\date{ \vspace*{0.5cm} This version: September, 2026}

\begin{document}
\bgroup

\begin{singlespace}
\maketitle

\begin{abstract}
\noindent
This paper combines synthetic aperture radar imagery, nighttime lights, and port characteristics to measure port-level maritime trade using only publicly available data. We evaluate the framework at ports the model has never observed, training on United States ports and testing on European ports and then reversing the exercise, and find that the mapping transfers in both directions. Applying the framework to Russian ports after the 2022 sanctions, we find reallocation, with the Far East the only basin rising faster than the movement common to all Russian ports. The approach complements methods based on the Automatic Identification System: where transponders broadcast, AIS-based estimates track official statistics closely, and after the sanctions the radar keeps recording activity at western Russian ports that disappears from the AIS record, which is where dark shipping concentrates.
\end{abstract}

\end{singlespace}
\vspace{0.25em}

\thispagestyle{empty}

\clearpage
\egroup
\setcounter{page}{1}

\section{Introduction\label{sec:intro}}

Satellite data have become a central tool for measuring economic activity where conventional statistics are delayed, incomplete, or unavailable. Beginning with the use of nighttime lights as proxies for income and growth \citep{elvidge1997relation, chen2011using, henderson2012measuring}, the economics literature has progressively expanded the range of remote-sensing applications \citep[see][for a survey]{donaldson2016view}. Around 80 percent of the volume of international trade in goods is carried by sea \citep{unctad2024}, yet port-level statistics are published by a minority of countries and are absent or unreliable exactly where they would be most useful. Sanctioned economies do not publish them, states with limited statistical capacity publish them late or not at all, and newly built terminals go unrecorded for years.

This paper asks whether publicly available imagery can measure a port's trade when no statistics for that port have ever existed. Alternative approaches based on the Automatic Identification System (AIS) are informative when transponder signals are available. For instance, \citet{arslanalp2025nowcasting} build a global nowcast of maritime trade from those signals, but any such approach is inherently limited when vessels intentionally disable transmission. \citet{fernandez2025charting} document this practice among oil tankers serving sanctioned economies and estimate that vessels sailing dark carry around two fifths of recorded seaborne crude. Our framework circumvents the limitation by relying on remote-sensing data and port characteristics that do not depend on voluntary transmission. \Cref{sub:dark} measures the difference directly: where transponders broadcast, the AIS record tracks official statistics almost exactly, and at western Russian ports after the sanctions it loses activity the radar sees.

The core intuition is that port activity leaves physical footprints visible from space. We combine three data sources. Synthetic aperture radar (SAR) imagery from Sentinel-1 records what moves between passes regardless of weather or daylight. Nighttime lights (NTL) from the Visible Infrared Imaging Radiometer Suite (VIIRS) proxy for continuous operation. The World Port Index (WPI) supplies 90 time-invariant characteristics of the port itself. We map these inputs to port-level trade with gradient-boosted trees \citep{chen2016xgboost}, and we evaluate the mapping by its performance at a port the model has never seen.

Our central result is that the mapping transfers, and that it transfers in both directions. Holding out one United States port at a time, predicted and actual port means correlate at 0.71 in rank. Training on the entire U.S. panel and applying the result to 113 European ports, with no European observation in training and no parameter taken from the target country, gives 0.58. Reversing the exercise and training on Europe to predict U.S. ports gives 0.62, and holding out one European port at a time gives 0.59. All four training and testing combinations lie between 0.58 and 0.71 and their bootstrap intervals overlap.

The imagery does work that the port characteristics do not. The characteristics alone rank held-out ports at 0.59 but reach a levels $R^2$ of only 0.067, and adding the satellite block raises that to 0.220. Two ports with the same recorded depth and the same terminal list can differ several-fold in throughput. The port characteristics carry the ordering; the imagery supplies the scale. Following \citet{meyer2021} we compute a dissimilarity index for each port and validate it out of sample, which turns a single accuracy figure for the whole panel into a per-port expected error and tells a reader which ports the model is entitled to speak about.

We apply the framework to Russia, whose ports have never published trade statistics and whose western terminals were sanctioned in February 2022. With port and month fixed effects, every basin rises after the break and only the Far East rises by more than the movement common to all Russian ports, with a difference-in-differences estimate of $+0.206$.\footnote{Following \citet{lee2018}, who studies redistribution within North Korea using luminosity, our claim is relative, not absolute.} The Russian application is also where the difference from AIS-based measurement becomes measurable. Where transponders broadcast, the AIS record is excellent: matched to official statistics, the IMF's AIS-based PortWatch correlates at 0.90 across U.S. ports, and at Russian ports the gap between the two records is flat through three pre-sanction years. After February 2022 the two records part (\Cref{fig:portwatch_gap}). The gap between the radar index and the AIS record opens by about 0.8 log points at Baltic and Black Sea ports and stays flat in the east, in every specification we report, and it opens exactly where \citet{fernandez2025charting} locate the dark fleet. The radar keeps seeing what the transponder stops reporting, so a measure relying on AIS alone could miss part of the western activity and overstate the eastward shift.

Our paper contributes to several literatures. First, it advances the satellite-based economic measurement agenda \citep{henderson2012measuring, donaldson2016view}, and in particular the strand that predicts output at individual facilities from geospatial features, as \citet{yang2025geospatial} do for steel plants, by extending it to port-level trade, a domain where conventional data are particularly prone to delay and suppression, and by carrying the cross-sectional half of that agenda to a spatial unit finer than any in the existing validation studies. Second, it complements AIS-based trade measurement \citep{arslanalp2025nowcasting, stamer2021, heiland2025, fernandez2025charting} with a framework that does not depend on voluntary transmission. Third, it contributes to the spatial machine learning literature. \citet{ploton2020spatial} show that random cross-validation overstates performance when observations are spatially structured, and \citet{meyer2021} give a way to say where a fitted model applies. We add transfer in both directions between two continents, and we report, port by port, where the transferred mapping is entitled to speak. Framed this way our measure belongs with the gridded output products of \citet{nordhaus2006}, which are coarse spatial measures adopted because no alternative exists.

The remainder of the paper proceeds as follows. \Cref{sec:data} describes the data sources and feature construction. \Cref{sec:ports} explains why every validation holds out ports and presents the transfer results. \Cref{sec:russia} applies the framework to Russian ports. \Cref{sec:conclusion} concludes.

\section{Data and Measurement \label{sec:data}}

\subsection{Unit of observation and area of interest}

Our unit of observation is a port-month, and we identify ports by their WPI number.\footnote{Two U.S. names each map to two distinct ports, Portland in Maine and Oregon and Wilmington in Delaware and North Carolina, so merging on name silently creates cross-port pairs.} Each port enters as a point, and we compute every satellite variable inside the bounding box of a 3-kilometer buffer around that point, a square covering about 28 square kilometers that is identical for every port. A fixed area of interest (AOI) is comparable across ports and requires no port-specific knowledge, exactly what a method for unmeasured ports needs, since berth outlines do not exist for the ports the measure is built for. \Cref{tab:aoi} in \Cref{app:data} shows that the ordering of ports survives at 2 and 4 kilometers.

\subsection{Synthetic aperture radar}

We use Sentinel-1 Ground Range Detected imagery in interferometric wide-swath mode, VV and VH polarization.\footnote{The pair names the transmit and receive orientation: VV transmits and receives vertically polarized signal, VH transmits vertically and receives horizontally. Metallic and vertical structures, vessels, cranes and containers among them, return the cross-polarized VH band strongly, so the two bands read a port differently.} Within a port-month, consecutive acquisitions are paired and the absolute difference is taken pixel by pixel and summed over the window,
\begin{equation}
D_{a,t} = \sum_{i,j} \left| I^{\text{VV}}_{a,t+1}(i,j) - I^{\text{VV}}_{a,t}(i,j) \right|,
\end{equation}
where $I^{\text{VV}}_{a,t}(i,j)$ is the pixel value at position $(i,j)$ of the window $a$ of acquisition $t$. The month's value is the median of those pair sums,\footnote{The median instead of the mean, for robustness to outliers from technical noise and scattering effects.}
\begin{equation}
VV_{a,m} = \operatorname{median}\left\{ D_{a,t} \mid t \in m \right\}.
\label{eq:sar_diff}
\end{equation}
The VH band serves as a stock proxy: for each port-month we take the pixel-wise median across the month's acquisitions and average it over the window,\footnote{VH backscatter is in decibels, so negative values are feasible and correspond to weak returns.}
\begin{equation}
VH_{a,m} = \frac{1}{|\Omega_a|} \sum_{(i,j)\in\Omega_a} \operatorname{median}_{t\in m} \left\{ I^{\text{VH}}_{a,t}(i,j) \right\},
\label{eq:vh_backscatter}
\end{equation}
with $\Omega_a$ the set of pixels in the window.

\begin{figure}[htbp]
  \centering
  \includegraphics[width=\textwidth]{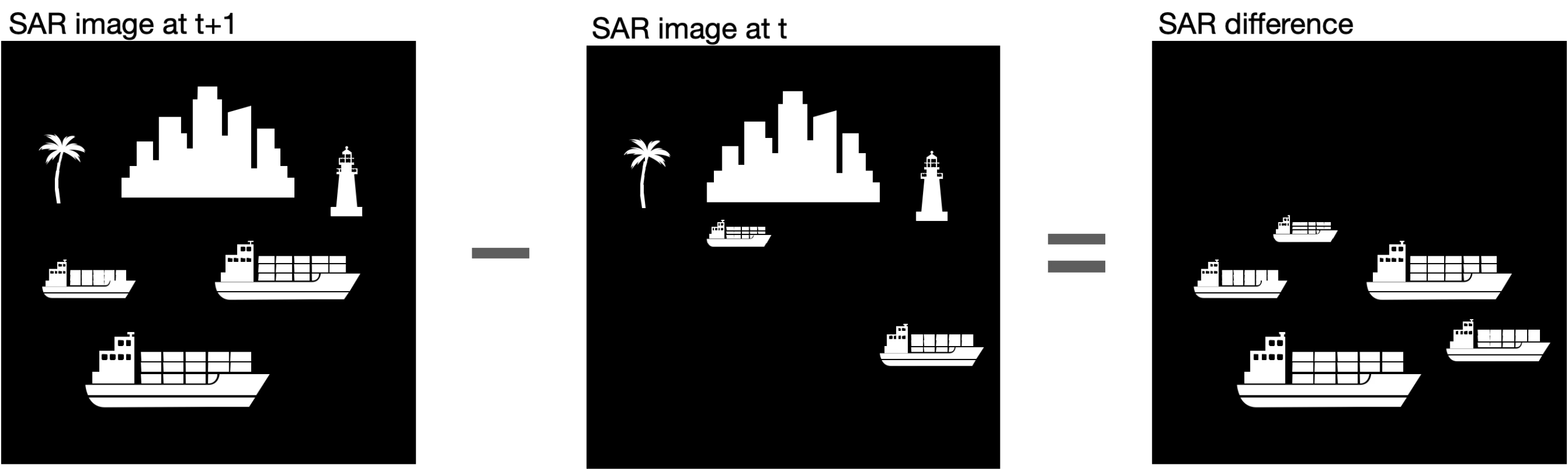}
  \caption{The image difference}
  \label{fig:sar_diff}
  \vspace{0.5em}
  \begin{minipage}{1\textwidth}
  \setstretch{1.15}
    \footnotesize
Notes: Illustration. Two consecutive radar acquisitions of the same window and
their pixel-wise difference. Structures present in both passes, the skyline
and the lighthouse here, drop out; what moved between passes, the vessels,
survives. The variable sums that difference over the window.
\end{minipage}
\end{figure}

\Cref{fig:sar_diff} illustrates the construction. What the variable measures is movement. Radar returns are unaffected by cloud, darkness and season, so the difference between two passes is dominated by what physically changed, which at a port means vessels berthing and departing, containers restacked and vehicles on the apron. A port where more moves between passes is a port doing more.

The variable carries information the port characteristics do not. Across 68 U.S. ports its port means span a factor of 563 and correlate 0.42 with total trade weight, imports and exports combined, or 0.55 in rank. Regressed on the seven numeric size fields of the WPI, it returns an $R^2$ of 0.162, so most of what it carries is absent from the WPI. \Cref{app:data} reports which pixels survive the decibel transform and how they divide between water and land, measured against the global surface water product of \citet{pekel2016}.

\subsection{Nighttime lights}

Daily observations from the VIIRS Black Marble product \citep{roman2018nasa} are aggregated over the same window to the monthly mean, sum, maximum and standard deviation of radiance, and to the share of the window above a fixed lit threshold,
\begin{equation}
\text{LR}_{a,m} = \frac{1}{|\Omega_a|} \sum_{(i,j)\in\Omega_a} \mathbf{1}\!\left\{ \operatorname{median}_{t\in m} \left[I^{\text{NTL}}_{a,t}(i,j)\right]>\tau\right\},
\label{eq:lar}
\end{equation}
with $\tau = 0.5$ following the Black Marble documentation. Since water is unlit at night, the lit share measures the spatial footprint of port infrastructure. A port working around the clock is lit and a dormant one is not. VIIRS is preferred to its predecessor because \citet{gibson2021} finds it predicts regional output substantially better. The calendar accounts for 53 percent of the within-port variation in mean radiance and 44 percent in its standard deviation, which is one further reason we use the measure cross-sectionally.

\subsection{Port characteristics}

Ninety fields of the WPI enter the model, covering harbor size and type, channel and berth depths, maximum vessel dimensions, terminal types, cranes and services. All are time-invariant. They describe what a port can handle and not what it does, which is why they fix the ordering of ports almost entirely and put almost no number on it.

\subsection{Trade data and coverage}

Our U.S. target is customs vessel weight and value by port-month from 2016 to 2024, which is 7,343 port-months over 68 ports. The European target is Eurostat gross weight handled at the quay, quarterly, for 113 ports matched to a WPI record. Russia contributes 35 ports and no target at all, which is the point of the exercise.

The two targets are different quantities. U.S. customs weight is cargo clearing customs on a vessel, while Eurostat gross weight is tonnage crossing the quay including domestic and transit cargo. A prediction trained on one and applied to the other misses the level badly, and \Cref{sec:ports} reports every cross-continent result as a statement about ordering alone.

\Cref{tab:coverage} reports panel coverage and the effect of the Sentinel-1B failure of December 2021. Coverage rises for the U.S. and European panels, because the 2016 ramp-up of the constellation costs more observations than the 2021 loss of one satellite. Only the Russian panel falls, from 0.867 to 0.675, two months before the event studied in \Cref{sec:russia}.

\subsection{Model}

We generate predictions with gradient-boosted regression trees, selecting hyperparameters by Bayesian search against a validation split defined over ports, since the object of the exercise is a port outside the training sample. The search for \Cref{tab:transfer} minimizes the root mean squared error of held-out port means, five port-grouped folds and 300 trials per training panel, and \Cref{tab:arms} states its own criterion where it differs. Two properties of that procedure are worth stating. The search is not nested inside the leave-one-port-out loop, so every scored port influenced the configuration, which is optimistic in level but applies identically to every comparison we make. And the categorical fields of the WPI are encoded once across all panels, so a category receives the same code everywhere; this uses no target, and encoding each panel separately would map categories seen only in Europe or Russia to an unseen code.

\Cref{tab:modelclass} shows that the choice of learner is immaterial. Nine model classes, each given its own 300-trial search under an identical protocol, span a narrow range, and the leading class is statistically distinguishable from only one of eight rivals on weight and two of eight on value. Ridge regression with a single hyperparameter falls inside the interval of the leader.

\section{Ports Without Statistics \label{sec:ports}}

The object of the exercise is a port that publishes no statistics, with Russian ports and ports in low-income countries the leading cases. What has to be established is that a mapping learned on ports that do report carries to a port that does not. Every validation in this paper therefore holds out ports: one U.S. port at a time, then one European port at a time, then an entire continent.

Pooled goodness-of-fit statistics computed at port-month level are dominated by differences between ports, and those differences are large. The between-port share of the variance of log trade is 0.86 for weight and 0.91 for value. A model that learns only which ports are large will report a high pooled $R^2$ while carrying no information about any port's own path, and \citet{gibsonboe2021} draw the same distinction for the nighttime-lights literature. Holding out ports is what makes the numbers below mean what they appear to mean. \Cref{app:results} reports the aggregate fit for completeness and shows how much of it is cross-sectional.

\subsection{The transfer matrix}

\Cref{tab:transfer} is our central result. Rows give the panel a model was trained on and columns the panel it was tested on. On the diagonal we hold out one port at a time. Off the diagonal an entire continent is unseen and contributes no observation to training. Features are identical in every combination, and we tune hyperparameters once per training panel and then apply them to both of that panel's test sets, which is the structure a practitioner faces when tuning on the data available and applying the result elsewhere.

\begin{table}[htbp]
\centering
\caption{Transfer of the port-scale mapping across continents}
\label{tab:transfer}
\begin{threeparttable}
\begin{tabular}{lcc}
\toprule
 & \multicolumn{2}{c}{Tested on} \\
\cmidrule(lr){2-3}
Trained on & U.S. & Europe \\
\midrule
U.S. & \makecell{0.710 \\ \footnotesize [0.560, 0.817]} & \makecell{0.581 \\ \footnotesize [0.439, 0.690]} \\[2pt]
Europe & \makecell{0.624 \\ \footnotesize [0.425, 0.777]} & \makecell{0.591 \\ \footnotesize [0.453, 0.698]} \\[2pt]
\midrule
\multicolumn{3}{l}{\textit{Two-parameter $R^2$ (scale and intercept fitted in the test panel)}} \\
U.S. & 0.410 & 0.350 \\
Europe & 0.316 & 0.286 \\
\bottomrule
\end{tabular}
\begin{tablenotes}[flushleft]\footnotesize
\item Notes: Spearman rank correlation between actual and predicted port means, with 95
percent intervals from 4,000 bootstrap draws over ports. On the diagonal a single
port is held out at a time; off the diagonal the entire test continent is unseen and
contributes no observation to training. Hyperparameters are tuned once per training
panel by 300 Optuna trials with five-fold grouping over that panel's ports, against
the root mean squared error of held-out port means, and the resulting configuration
is applied to both of its test sets. Features are identical
in every combination: eleven satellite variables and ninety WPI fields.
Because U.S. customs weight and Eurostat quay tonnage are different quantities on
different scales, the raw levels $R^2$ off the diagonal is -80 and -32; the mapping transfers in rank, not in level.
\end{tablenotes}
\end{threeparttable}
\end{table}

\begin{figure}[htbp!]
  \centering
  \begin{subfigure}[t]{0.48\textwidth}
    \centering
    \includegraphics[width=\textwidth]{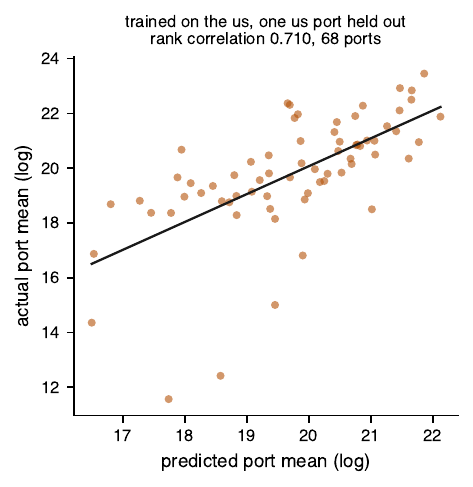}
    \caption{U.S.\ to U.S.}\label{fig:transfer_us_us}
  \end{subfigure}
  \hfill
  \begin{subfigure}[t]{0.48\textwidth}
    \centering
    \includegraphics[width=\textwidth]{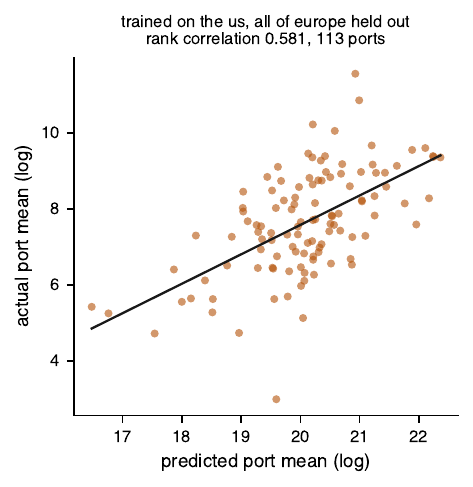}
    \caption{U.S.\ to Europe}\label{fig:transfer_us_eu}
  \end{subfigure}
  \\[0.8em]
  \begin{subfigure}[t]{0.48\textwidth}
    \centering
    \includegraphics[width=\textwidth]{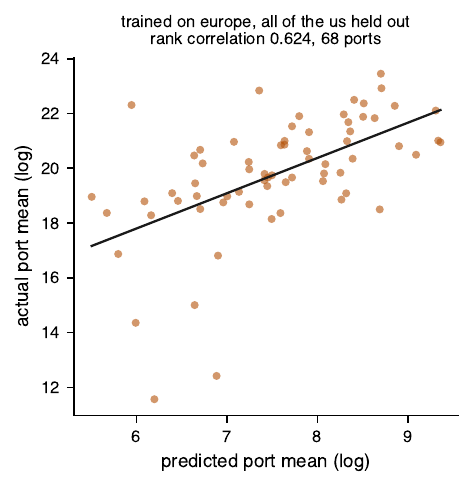}
    \caption{Europe to U.S.}\label{fig:transfer_eu_us}
  \end{subfigure}
  \hfill
  \begin{subfigure}[t]{0.48\textwidth}
    \centering
    \includegraphics[width=\textwidth]{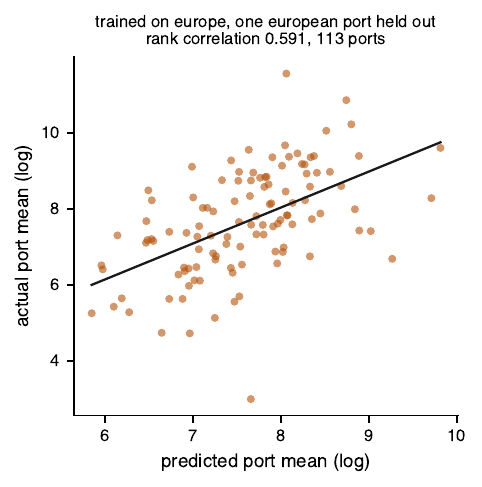}
    \caption{Europe to Europe}\label{fig:transfer_eu_eu}
  \end{subfigure}
  \caption{Actual against predicted port means in all four training and testing
  combinations}
  \label{fig:transfer}
  \vspace{0.5em}
  \begin{minipage}{1\textwidth}
  \setstretch{1.15}
    \footnotesize
Notes: Panels (a) and (d) hold out one port at a time inside a panel. Panels (b) and
(c) train on one continent and test on the other, with no observation from the test
continent in training. Each point is a port mean over the sample period. Axes are in
logs of each panel's own target, which are different quantities, so the fitted line
is a slope and not the 45-degree line.
\end{minipage}
\end{figure}

Three readings follow. \Cref{fig:transfer} plots the four combinations. Every one lies between 0.58 and 0.71 and all four intervals overlap, so the advantage of training on the continent one intends to predict is not resolvable at this number of ports. The reverse direction works, since a model that has seen only European ports ranks U.S. ports at 0.62, which means the mapping is not an artifact of the U.S. panel. Third, the level does not transfer at all, as \Cref{tab:levelgap} shows. A single additive shift, which uses no port-specific information, removes almost all of the raw squared error off the diagonal and takes the figure to $+0.320$ and $+0.300$. The further rescaling removes almost nothing. On the diagonal, where the two targets are the same quantity, there is nothing for either step to remove. The cross-continent figure is a difference of units and not a failure of ordering, and it is why we report an ordering, not a quantity.

Withdrawing an entire U.S. region instead of a single port costs almost nothing: the twenty-one East Coast ports held out together come back at 0.719 against 0.710 one at a time, so whatever the model has learned is not a memory of the particular ports it saw (\Cref{tab:region} in \Cref{app:results}).

\subsection{The satellite contribution}

\begin{table}[htbp]
\centering
\caption{The satellite contribution}
\label{tab:contribution}
\begin{threeparttable}
\begin{tabular}{lcccc}
\toprule
Specification & Ports & Rank correlation & 95\% interval & Levels $R^2$ \\
\midrule
Characteristics only & 66 & 0.593 & [0.366, 0.749] & 0.067 \\
Satellite only & 68 & 0.516 & [0.317, 0.681] & 0.131 \\
Satellite $+$ characteristics & 66 & 0.668 & [0.507, 0.781] & 0.220 \\
\bottomrule
\end{tabular}
\begin{tablenotes}[flushleft]\footnotesize
\item Notes: Leave-one-port-out over U.S. ports, weight target. The characteristics carry the
ordering; the imagery supplies the scale. A model that knows nothing about an unseen
port predicts a constant, which carries no ordering information.
\end{tablenotes}
\end{threeparttable}
\end{table}

\Cref{tab:contribution} separates the port characteristics from the imagery: the characteristics alone rank held-out ports at 0.593, most of the ordering, with a levels $R^2$ of 0.067, and the satellite block raises that to 0.220. Depth and terminal counts say which ports are capable of volume; the imagery says how much is actually happening.

\subsection{Direction of trade}

\begin{figure}[htbp!]
  \centering
  \includegraphics[width=0.78\textwidth]{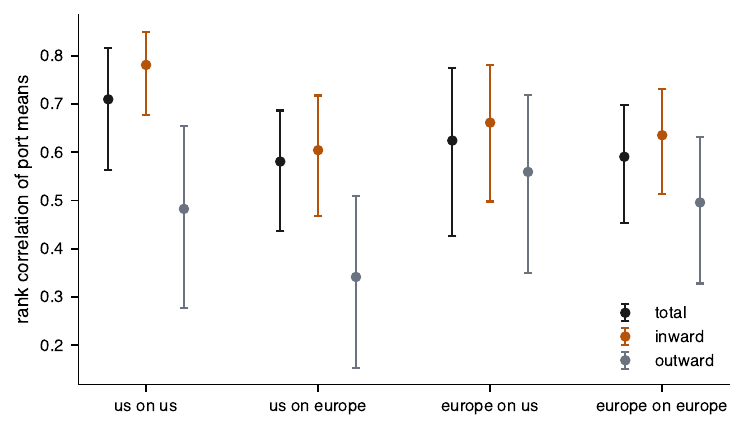}
  \caption{Transfer by direction of trade}
  \label{fig:direction}
  \vspace{0.5em}
  \begin{minipage}{1\textwidth}
  \setstretch{1.15}
    \footnotesize
Notes: Rank correlation of port means in each training and testing
combination, by direction of trade, with 95 percent port-bootstrap intervals.
\end{minipage}
\end{figure}

The two panels count different things, and one difference can be removed without new data. U.S. customs data report import and export vessel weight as separate series whose sum is the total, and the Eurostat series carries a direction dimension. Neither split is estimated. \Cref{tab:direction} runs the whole matrix three times, once on each. Arriving tonnage is better predicted than the total in every combination, and departing tonnage is worse in every one. Holding out one U.S. port, the rank correlation rises from 0.710 on the total to 0.781 on arrivals and falls to 0.483 on departures, and the same ordering holds across continents and inside Europe. \Cref{fig:direction} shows the twelve cells together.

The asymmetry has a physical reading. Cargo arriving is discharged and dwells in the terminal, in stacks and yards, until it is collected. Cargo leaving arrives from inland and is loaded, so it passes through the window quickly. What the radar sees is what is present when the satellite passes, and arrivals are present for longer. The measure is therefore better at what comes in than at what goes out, and we say so where it matters, in \Cref{sec:russia}.

Matching the direction does not close the cross-continent gap, however. The mean across the two cross-continent combinations rises from 0.602 to 0.633, a change small against intervals of about 0.25. The concept difference between the two targets is part of what separates the continents and it is not the binding part, which is consistent with \Cref{tab:levelgap}.

\subsection{Cargo composition \label{sub:cargo}}

\begin{table}[htbp]
\centering
\caption{Ordering accuracy by measured cargo composition, European ports}
\label{tab:cargo_observed}
\begin{threeparttable}
\begin{tabular}{lcccc}
\toprule
& \multicolumn{2}{c}{Trained on the U.S.} &
  \multicolumn{2}{c}{Trained on Europe} \\
\cmidrule(lr){2-3}\cmidrule(lr){4-5}
& Ports & Rank correlation & Ports & Rank correlation \\
\midrule
\multicolumn{5}{l}{\emph{Panel A. Total tonnage, ports grouped by their dominant
cargo class}} \\
All European ports & 113 & \makecell{0.581\\\footnotesize [+0.436, +0.686]} & 113 & \makecell{0.591\\\footnotesize [+0.453, +0.697]} \\
\midrule
Liquid bulk & 27 & \makecell{0.608\\\footnotesize [+0.311, +0.805]} & 27 & \makecell{0.433\\\footnotesize [+0.002, +0.741]} \\
Dry bulk & 42 & \makecell{0.531\\\footnotesize [+0.276, +0.715]} & 42 & \makecell{0.543\\\footnotesize [+0.293, +0.727]} \\
Containers & 17 & \makecell{0.289\\\footnotesize [-0.162, +0.630]} & 17 & \makecell{0.429\\\footnotesize [-0.098, +0.768]} \\
Roll-on, self-propelled & 12 & \makecell{0.510\\\footnotesize [-0.124, +0.921]} & 12 & \makecell{0.573\\\footnotesize [-0.033, +0.922]} \\
Roll-on, other & 8 & \makecell{0.429\\\footnotesize [-0.315, +1.000]} & 8 & \makecell{0.286\\\footnotesize [-0.595, +0.918]} \\
Other cargo & 7 & \makecell{0.786\\\footnotesize [+0.059, +1.000]} & 7 & \makecell{0.429\\\footnotesize [-0.623, +1.000]} \\
\midrule
\multicolumn{5}{l}{\emph{Panel B. Tonnage of one cargo class at a time, all ports
reporting it}} \\
\midrule
Liquid bulk & 109 & \makecell{0.503\\\footnotesize [+0.340, +0.633]} & 104 & \makecell{0.475\\\footnotesize [+0.298, +0.625]} \\
Dry bulk & 112 & \makecell{0.310\\\footnotesize [+0.128, +0.475]} & 112 & \makecell{0.270\\\footnotesize [+0.077, +0.447]} \\
Containers & 101 & \makecell{0.436\\\footnotesize [+0.268, +0.583]} & 97 & \makecell{0.413\\\footnotesize [+0.221, +0.577]} \\
Roll-on, self-propelled & 96 & \makecell{0.182\\\footnotesize [-0.022, +0.374]} & 86 & \makecell{0.181\\\footnotesize [-0.028, +0.376]} \\
Roll-on, other & 87 & \makecell{0.240\\\footnotesize [+0.038, +0.430]} & 73 & \makecell{0.050\\\footnotesize [-0.171, +0.276]} \\
Other cargo & 112 & \makecell{0.140\\\footnotesize [-0.064, +0.327]} & 110 & \makecell{0.064\\\footnotesize [-0.129, +0.247]} \\
\bottomrule
\end{tabular}
\begin{tablenotes}[flushleft]\footnotesize
\item Notes: Eurostat reports the tonnage that forms the European target split into six
disjoint cargo classes. The six sum to the reported total within one percent in 99.6
percent of port-quarters. A port's dominant class is the one with the largest share of
its tonnage over the whole sample. Panel A regroups the held-out predictions of
\Cref{tab:transfer} and refits nothing. Panel B changes the target, so each class is
tuned on its own with 300 trials under five-fold grouped cross-validation over ports,
and the configurations are reported in the replication package. The U.S.
column trains on U.S. ports alone and holds the whole European panel out.
The European column leaves one European port out at a time. Intervals are 4,000
bootstrap draws over ports.
\end{tablenotes}
\end{threeparttable}
\end{table}

Direction is one way the two panels differ. What is being moved is another, and it matters for what follows, because Russian export terminals handle mostly liquid bulk. Eurostat reports the European tonnage split into six disjoint cargo classes, so in Europe the composition can be measured, not assumed. The six sum to the reported total within one percent in 99.6 percent of port-quarters. \Cref{tab:cargo_observed} asks two questions of it. Panel A keeps the target at total tonnage and groups the held-out European ports by the class that accounts for most of their tonnage, refitting nothing. Panel B changes the target to the tonnage of one class at a time, which is a different problem, so each class is tuned on its own.

The ordering of total tonnage survives conditioning on port type. Trained on U.S. ports alone, it reaches 0.608 at the 27 European ports that are mostly liquid bulk and 0.531 at the 42 that are mostly dry bulk, against 0.581 across all 113. The absolute error is unrelated to a port's liquid-bulk share, with a rank correlation of $0.000$ across the 110 ports where both are observed. Nothing about the result is carried by one kind of port.

Asked for one class at a time the model does worse than it does on the total, and it should. The imagery records that a berth is occupied without recording what crosses it, so a change detector over a whole terminal reads the sum better than any part of it. The classes then separate in the way that reading predicts. Liquid bulk at 0.503 and containers at 0.436 are handled at dedicated infrastructure with a fixed footprint. Roll-on traffic at 0.182 and the residual class at 0.140 are not, and both fall inside the interval of no ordering. Liquid bulk is the best ordered of the six classes when a single class is the target. That is the class \Cref{sec:russia} depends on, and it is the one the imagery reads best.

\subsection{Resolution}

The measure resolves scale and stops short of fine differences. \Cref{tab:resolution} reports the share of port pairs whose ordering is called correctly, binned by how far apart the two ports actually are. For pairs differing by more than twenty-fold the ordering is right 88 percent of the time, and for pairs within a factor of 1.6 it is right 53 percent of the time, which is chance. Within size terciles the figure runs from 53 to 61 percent. This bounds the questions the measure can be asked, and it directly answers the question of which ports are predicted well and which are not.

\subsection{Learning curve \label{sub:learning}}

\begin{figure}[htbp]
  \centering
  \includegraphics[width=0.72\textwidth]{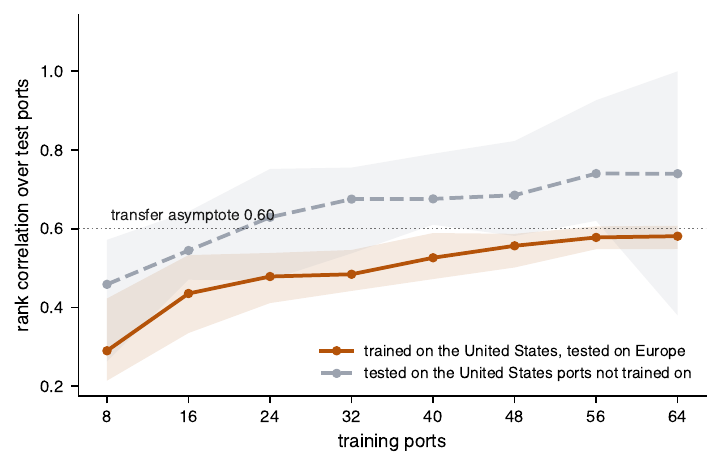}
  \caption{The learning curve}
  \label{fig:learning}
  \vspace{0.5em}
  \begin{minipage}{1\textwidth}
  \setstretch{1.15}
    \footnotesize
Notes: Training ports drawn at random without replacement, 20 draws at each size, the
same seeds at every size so the curves are paired. Bands are the tenth to ninetieth
percentile across draws. The configuration of \Cref{tab:transfer} is held fixed at every size, so the
curve is the curve for that model and small samples are not retuned in their own
favor. The upper series tests on the U.S. ports each draw did not train on,
so its test set shrinks as the training set grows and its band widens accordingly, with
four ports left at the largest size. The lower series tests on all 113 European ports,
which are not in the training sample at any size, and is the comparison the
paper's claim rests on. The dotted line is the asymptote of a curve $\rho(K) = a - b/K$ fitted to the eight
means of that series.
\end{minipage}
\end{figure}

A measure that transfers at 0.58 invites the question of whether it would transfer better with more labeled ports. \Cref{fig:learning} answers it by training on random subsets of the U.S. panel and testing on the whole of Europe, which is not in the training sample at any size.

The curve rises steeply and then stops. Eight training ports give 0.290 and sixty-four give 0.581, but the last eight of those buy 0.003. Fitting $\rho(K) = a - b/K$ to the eight means puts the asymptote at 0.600, so the full panel is within two hundredths of what an unlimited number of U.S. ports would deliver, and doubling the panel to 136 would buy about 0.01 more.

What limits the measure is therefore not how many ports were labeled. The configuration is held fixed across sizes, which is generous to the large end, so the plateau is if anything overstated in the flattering direction and the conclusion does not turn on it.

That leaves the features and the target, and the target can be checked. A port's average is taken over a median of 98 months, so its sampling error is small against the spread between ports, 0.028 against 2.188 log points. Three routes agree on how small. A components calculation puts the reliability of the port mean at 0.998, an odd-against-even split corrected to full length puts it at 0.999, and simulating two noisy copies of the same truth puts the implied ceiling on a rank correlation at 0.998. The European panel gives 0.999 on all three. So a model that recovered the true ordering exactly would score essentially one, and the held-out 0.668 and the transferred 0.581 sit at two thirds and three fifths of that ceiling.\footnote{Throughout, 0.668 is the held-out rank correlation on the 66-port common sample of \Cref{tab:contribution}, where every specification has observations; the 68-port diagonal of \Cref{tab:transfer} puts the same design at 0.710. The exercise is the same and the sample differs.} The distance is not noise in the label.

One thing this cannot see. It bounds the sampling part of the target's error but says nothing about what customs itself records, since customs is the only source, so the true ceiling is lower by an unknown amount. The direction survives that. What stands between the measure and the ordering it is scored against is the description of the image, neither the number of labels nor the precision of the ones we have.

\subsection{Applicability}

\begin{figure}[htbp!]
  \centering
  \includegraphics[width=0.72\textwidth]{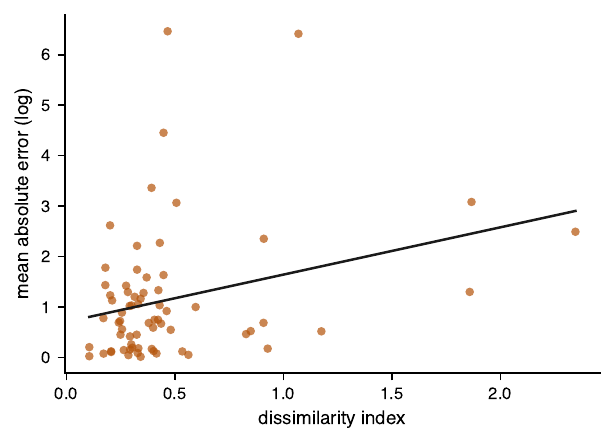}
  \caption{Held-out error against the dissimilarity index}
  \label{fig:di}
  \vspace{0.5em}
  \begin{minipage}{1\textwidth}
  \setstretch{1.15}
    \footnotesize
Notes: One point per U.S.\ port. The index is computed without using the
port's own outcome, so the relation is what licenses a per-port statement of
expected accuracy for ports with no statistics at all.
\end{minipage}
\end{figure}

A single accuracy figure for a whole panel is of little use to someone who wants to know about one port. Following \citet{meyer2021} we compute a dissimilarity index for each port, defined as that port's distance in predictor space from the training data with predictors weighted by their importance in the fitted model, and we validate it out of sample. Across 60 random half-splits of the U.S. panel it predicts held-out error with an average rank correlation of $+0.251$, positive in 56 of 60 splits, and on the full panel the correlation is $+0.295$ with a $p = 0.015$. \Cref{fig:di} plots that relation and \Cref{tab:applicability} collects it. The median held-out mean absolute error of the primary specification is 1.07 log points, a factor of 2.9 at the median port, and the index says which ports sit above and below that. The index then places the Russian ports. None of the 35 exceeds the maximum dissimilarity observed inside the U.S. training data, so none asks the model to extrapolate beyond the range it was fitted on, but 33 of them sit in the far quarter of that range, and we write \Cref{sec:russia} accordingly.

\section{Russian Ports After February 2022 \label{sec:russia}}

\subsection{Design}

Russia publishes no port-level trade statistics, and the February 2022 sanctions fell unevenly across its coasts, since Baltic and Black Sea terminals face European buyers while Far Eastern terminals face Asian ones. The sanctions are among the most extensive imposed on a major economy \citep{morgan2023economic, itskhoki2024economics, egorov2025trade}, and the literature measuring their trade effect works from customs and transaction records, on the 2014 round \citep{crozet2020friendly, crozet2021worth}, on the 2022 round through partner and Russian filings \citep{flach2024quantifying, babina2023assessing, de2026reducing}, and through the third countries that reroute sanctioned goods \citep{tyazhelnikov2024russian, chupilkin2023eurasian, li2024comply, chupilkin2026redflags, scheckenhofer2026evading}. None of those records names a Russian port, and the port is the resolution this section supplies.

What licenses this application is \Cref{sec:ports}. A model trained on one continent ranks the ports of another it has never seen, and it does so in both directions, so the mapping is a property of ports and imagery, not of the panel it was fitted on. Russia is the case the method was built for, and \Cref{tab:applicability} states how far it sits from the fitted range. These terminals handle mostly liquid bulk, and \Cref{sub:cargo} shows that this is the class the measure reads best, both when liquid-bulk ports are ordered on their total tonnage and when liquid-bulk tonnage itself is the target. The case this section rests on is not the awkward one for the method.

Estimating on a predicted outcome is unbiased so long as the prediction error does not move with the treatment, and it costs precision, a standard deviation 2.8 times what an outcome without prediction error would give; \Cref{tab:montecarlo} in \Cref{app:results} reports the simulation, calibrated from the held-out U.S. exercise. One case does bias it, and it is the assumption this section rests on. If the model's error shifts differently in the east after the break, the estimate moves one for one with that shift. Nothing inside the design can rule this out. What can be said against it is that the model is fitted on ports in other continents and sees no Russian observation, so there is no mechanism by which February 2022 changes its behavior at eastern ports and not at western ones, and that running \Cref{eq:did} at placebo break dates, February of 2018 through 2021 on the sample ending before the real break, gives estimates between $-0.029$ and $+0.017$ with permutation $p$ above 0.62 at every date (\Cref{tab:placebo_est}). No systematic error drifts east at arbitrary dates.

The design follows \citet{lee2018}, who studies redistribution within North Korea using luminosity. Our outcome is the model's predicted log trade and our claim is relative. At each date the model orders the 35 Russian ports by predicted trade, and our question is whether that ordering shifted after the break:
\begin{equation}
\hat{y}_{pt} = \alpha_p + \delta_t + \beta \, E_p \times \text{Post}_t + \varepsilon_{pt},
\label{eq:did}
\end{equation}
where $\hat{y}_{pt}$ is predicted log trade at port $p$ in month $t$, $\alpha_p$ and $\delta_t$ are port and month effects, $E_p$ marks the Far Eastern basin, and $\text{Post}_t$ switches on in February 2022. The continuous design replaces $E_p$ with standardized longitude. Inference reassigns the basin labels across ports, 3,000 draws, so the null is that the east--west assignment is exchangeable across ports. With port and month effects the accuracy of the level does not enter, and the shift is measured against the movement common to all Russian ports. The port effects also absorb each port's own ice regime, and the freeze-up and breakup cycles that differ across basins are seasonal, the margin the basin-by-calendar-month specification of \Cref{tab:embargo} and the winter placebo of \Cref{app:results} address directly.

Monthly ordering is what the design needs, and the model supplies it. On held-out U.S. ports, the cross-sectional rank correlation computed separately in each of 108 months averages 0.624 and falls below 0.5 in two of them, against 0.668 for the ordering of period means. Ranking a month is barely harder than ranking the average. Out of domain it is 0.539 over 36 quarters of the European panel from a model that has seen no European port, below 0.5 in one of them, and it does not decay after the Sentinel-1B failure or inside the test block. \Cref{fig:perdate} shows both.

\subsection{Reallocation, not contraction}

\begin{table}[htbp]
\centering
\caption{Russian ports after February 2022}
\label{tab:russia}
\begin{threeparttable}
\begin{tabular}{lccc}
\toprule
Basin & Ports & Change & Relative to the common change \\
\midrule
Far East & 12 & +0.262 & +0.116 \\
Azov-Black Sea & 4 & +0.107 & -0.039 \\
Arctic & 10 & +0.106 & -0.040 \\
Baltic & 9 & +0.064 & -0.082 \\
\midrule
All Russian ports & 35 & +0.146 & --- \\
\bottomrule
\end{tabular}
\begin{tablenotes}[flushleft]\footnotesize
\item Notes: Outcome is the model's predicted log trade, standardized, from the
specification trained on the U.S. panel with the customs weight target and
tuned as \Cref{tab:arms} describes. ``Change'' is the basin's own before-and-after
difference with port effects removed, and the final column subtracts the change
common to all Russian ports, both computed with equal weight on ports. The
difference-in-differences coefficient below weights port-months on the unbalanced
panel and contrasts the Far East with the other basins instead of the all-port
mean, so it is larger than the final column by construction. The difference-in-differences estimate for the Far East interacted with the
post-sanction indicator is +0.206 with a port-permutation $p = 0.100$, and with longitude as a continuous
exposure it is +0.097, $p = 0.105$. \Cref{tab:arms} reports every specification and \Cref{tab:power}
what this design could have detected.
\end{tablenotes}
\end{threeparttable}
\end{table}

\begin{figure}[htbp!]
  \centering
  \begin{subfigure}[t]{0.48\textwidth}
    \centering
    \includegraphics[width=\textwidth]{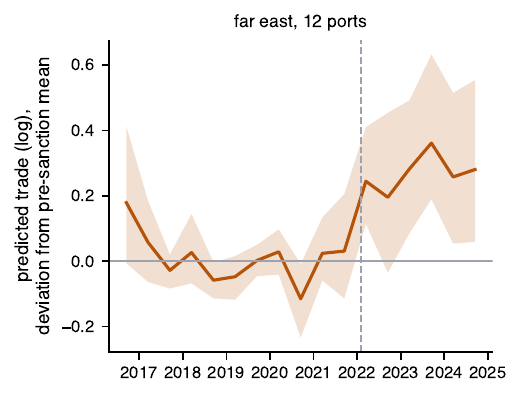}
    \caption{Far East}\label{fig:event_far_east}
  \end{subfigure}
  \hfill
  \begin{subfigure}[t]{0.48\textwidth}
    \centering
    \includegraphics[width=\textwidth]{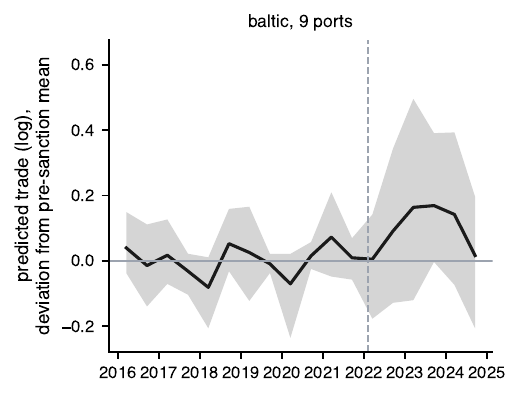}
    \caption{Baltic}\label{fig:event_baltic}
  \end{subfigure}
  \\[0.8em]
  \begin{subfigure}[t]{0.48\textwidth}
    \centering
    \includegraphics[width=\textwidth]{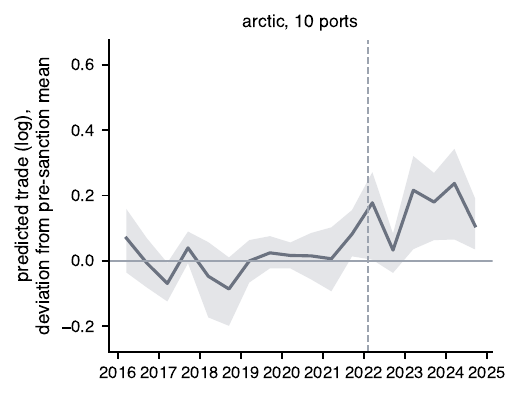}
    \caption{Arctic}\label{fig:event_arctic}
  \end{subfigure}
  \hfill
  \begin{subfigure}[t]{0.48\textwidth}
    \centering
    \includegraphics[width=\textwidth]{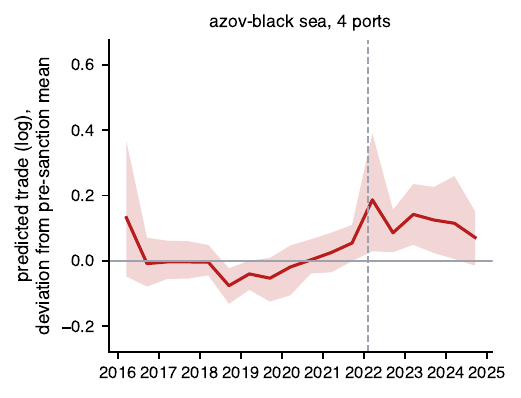}
    \caption{Azov--Black Sea}\label{fig:event_azov}
  \end{subfigure}
  \caption{Predicted trade at Russian ports by basin, half-yearly}
  \label{fig:russia_event}
  \vspace{0.5em}
  \begin{minipage}{1\textwidth}
  \setstretch{1.15}
    \footnotesize
Notes: Each panel plots the mean deviation of predicted log trade from each port's
own pre-sanction mean, averaged over the ports of that basin. Bands are 95 percent
intervals from 2,000 bootstrap draws over those ports. A half-year is plotted when at
least three ports of the basin are observed in it. The dashed line marks February
2022. All four panels share a vertical scale.
\end{minipage}
\end{figure}

Every basin rises after the break. The movement common to all Russian ports is $+0.146$ standard deviations and only the Far East rises by more. Reporting the fixed-effects coefficient alone, as \Cref{tab:russia} makes clear, would read as a decline in three basins, which is not what the data say, so \Cref{tab:russia} reports the absolute change and the relative change side by side. The difference-in-differences estimate for the Far East interacted with the post-sanction indicator is $+0.206$ with a port-permutation $p = 0.100$, and with longitude as a continuous exposure, which uses the whole east to west gradient, not a binary split, it is $+0.097$ with a $p = 0.105$. That permutation reassigns the east and west labels across ports and refits, so it asks whether the split we observe in the predictions is larger than one an arbitrary line would produce.\footnote{Two of the eleven satellite features are acquisition counts, and the radar count shifts east against west when Sentinel-1B fails. Refitting the primary arm from its recorded configuration and predicting with each port's counts frozen at their pre-break medians leaves the estimate at $+0.206$ with a permutation $p = 0.086$, so the shift in observation counts contributes nothing to the result.} The accuracy of the mapping behind those predictions is established in \Cref{sec:ports}, and \Cref{sub:corroboration} checks the direction against independent sources. Of the 71 monthly pre-period contrasts between eastern and western ports, two are significant at five percent, about the rate chance produces, so parallel trends are not contradicted, which is weaker than saying they are verified and is what our evidence supports. The replication package reports every lead.

Radar coverage of the Russian panel falls after the break, from 0.867 to 0.675 of port-months, and the loss concentrates in the Arctic, where seven of ten ports drop out. Port effects compare each port with itself, so a port absent after the break contributes nothing to the post-break comparison, and restricting to the 24 ports observed on both sides moves the direction estimates from $+0.050$ to $+0.049$ on total tonnage and from $+0.183$ to $+0.180$ on arrivals, and the primary estimate from $+0.206$ to $+0.205$.

\begin{table}[htbp]
\centering
\caption{The Far East estimate across every specification}
\label{tab:arms}
\begin{threeparttable}
\begin{tabular}{lllccc}
\toprule
Trained on & Target & Transform & Monthly rank & Far East & Permutation $p$ \\
\midrule
U.S. & Value & month-centered & 0.771 & -0.006 & 0.909 \\
U.S. & Value & as reported & 0.761 & +0.114 & 0.113 \\
U.S. & Weight & month-centered & 0.738 & +0.187 & 0.133 \\
U.S. & Weight & as reported & 0.736 & +0.206 & 0.100 \\
Both & Weight & month-centered & 0.636 & +0.059 & 0.472 \\
Both & Weight & as reported & 0.620 & +0.059 & 0.389 \\
Europe & Weight & month-centered & 0.520 & +0.006 & 0.878 \\
Europe & Weight & as reported & 0.561 & +0.000 & 0.996 \\
\bottomrule
\end{tabular}
\begin{tablenotes}[flushleft]\footnotesize
\item Notes: Every specification uses the paper's feature set, the eleven satellite
variables with the ninety WPI fields, and each receives its own 300-trial
search maximizing the mean over months of the cross-sectional rank correlation at
held-out ports, which is the column headed ``Monthly rank''. That criterion is
computed on the training panels with no Russian observation, and no specification was
selected. The value target exists only for the U.S., since the European
target is quay tonnage. Seven of the eight
estimates are positive and the eighth is $-0.006$. The spread is driven by the
training panel, not by the target or the transform, and the European arms,
whose predictions have the least within-port movement, return a null.
\end{tablenotes}
\end{threeparttable}
\end{table}

That estimate comes from one specification, so we report every specification we can defend. \Cref{tab:arms} varies the training panel, the target, and whether the target is centered on its own month, holding the feature set at the paper's. Each arm receives its own search against the monthly ranking criterion, computed on the training panels with no Russian observation, and we select none of them. Seven of the eight estimates are positive, the smallest at $+0.0002$, and the eighth is $-0.006$. The spread is driven by the training panel. The three positive U.S.-trained arms give $+0.206$, $+0.187$ and $+0.114$ with permutation $p = 0.100$, 0.133 and 0.113, the two pooled arms give about $+0.06$, and the two trained on Europe give essentially zero. Those European arms also carry the least within-port movement in their predictions, 1.3 and 2.4 percent against 6.1 percent for the primary arm, so a model that barely moves returns a null, which is what one would expect of it, not independent evidence against the result. The criterion column of \Cref{tab:arms} is diagnostic and not a selector. The reporting rule naming the primary arm was fixed before any Russian estimate was computed, on three ex ante grounds. Weight is the physical quantity a reallocation question is about. A value-trained model imports U.S. price and quantity relationships that the post-2022 Urals discount breaks at exactly the ports under study. And no value target exists for Europe, so weight is the only target the European arms can share.

\subsection{Dark shipping \label{sub:dark}}

\begin{table}[htbp]
\centering
\caption{The AIS record beside the radar index}
\label{tab:portwatch_cal}
\begin{threeparttable}
\begin{tabular}{lcc}
\toprule
 & Rank correlation & $n$ \\
\midrule
\multicolumn{3}{l}{\textit{Panel A. U.S. ports, monthly, 2019--2024}} \\
official statistics and PortWatch & 0.901 & 60 ports \\
radar index and official statistics & 0.661 & 60 \\
radar index and PortWatch & 0.602 & 60 \\[2pt]
ordering of ports at each month, PortWatch & 0.909 & 72 months \\
ordering of ports at each month, radar index & 0.585 & 72 \\
\midrule
\multicolumn{3}{l}{\textit{Panel B. European ports, quarterly}} \\
official statistics and PortWatch & 0.874 & 105 ports \\
radar index and PortWatch & 0.481 & 105 \\
\midrule
\multicolumn{3}{l}{\textit{Panel C. Russian ports, 2019-01 to 2022-01}} \\
radar index and PortWatch, port calls & 0.458 & 23 ports \\
radar index and PortWatch, tonnage & 0.361 & 23 \\
\bottomrule
\end{tabular}
\begin{tablenotes}[flushleft]\footnotesize
\item Notes: Spearman rank correlations of port means over the stated window; the
per-month rows average the cross-port rank correlation over months with at least
fifteen matched ports. PortWatch \citep{arslanalp2025nowcasting} estimates daily
port-level trade volumes from satellite AIS; 286 of the 318 ports in the three
panels match a PortWatch port within 25 km. The radar index is the paper's
specification unchanged: held-out predictions in Panel A, the U.S.-trained transfer in Panel B, and the primary Russian arm in Panel C.
\end{tablenotes}
\end{threeparttable}
\end{table}

Would a different instrument have seen the same thing? The natural alternative at this unit of observation is the Automatic Identification System. PortWatch \citep{arslanalp2025nowcasting} publishes daily AIS-based estimates of port calls and trade volumes for 2,065 ports, and 286 of the 318 ports in this paper's three panels match one within 25 kilometers. \Cref{tab:portwatch_cal} says how good that record is where it can be checked. Against official statistics the PortWatch port means correlate at 0.901 across U.S. ports and 0.874 across European ports, and it orders U.S. ports month by month at 0.909 over the 72 common months from 2019, where this paper's index reaches 0.585. Where transponders broadcast, the AIS record is close to official statistics at exactly the unit this paper measures, and nothing in this paper improves on it there.

\begin{figure}[htbp]
  \centering
  \includegraphics[width=0.85\textwidth]{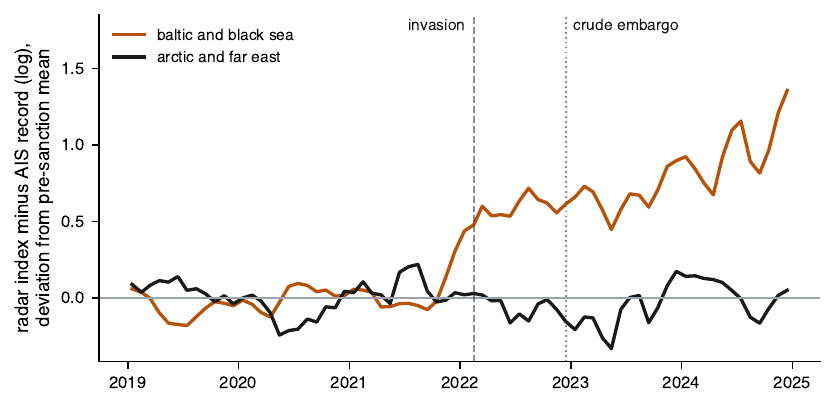}
  \caption{The radar--AIS gap at Russian ports}
  \label{fig:portwatch_gap}
  \vspace{0.5em}
  \begin{minipage}{1\textwidth}
  \setstretch{1.15}
    \footnotesize
Notes: The gap is the paper's predicted log trade minus the log of PortWatch
estimated tonnage at the same port and month, shown as the deviation from each
port's pre-sanction mean and averaged within each group of basins, smoothed with a
three-month centred moving average. The dashed line marks February 2022 and the
dotted line the December 2022 crude embargo, when \citet{fernandez2025charting}
document the jump in dark shipping. The monthly values behind the figure are in the
replication package; \Cref{tab:portwatch_gap} reports the estimates and their
permutation inference.
\end{minipage}
\end{figure}

The case it cannot cover is the one \Cref{sec:russia} is about. \citet{fernandez2025charting} estimate that vessels sailing dark carry around two fifths of recorded seaborne crude, that the dark share jumped when the price cap took effect, and that the activity concentrates at Baltic and Black Sea ports. The radar does not depend on the transponder, so the two records can be compared at the same port and month. In the pre-sanction years they agree: across the 23 Russian ports with both records, the port means correlate at 0.458 on port calls ($p = 0.028$). \Cref{tab:portwatch_gap} then asks whether the gap between them, predicted log trade minus the logged AIS record, with port and month effects, opened in the west after February 2022. It did, in all eight specifications we report: by 0.786 log points against AIS tonnage (permutation $p = 0.063$) and 0.727 against port calls ($p = 0.040$), unchanged by a west-by-calendar-month seasonal, and largest after the December 2022 embargo, which is when the dark share jumped. \Cref{fig:portwatch_gap} shows the gap flat in both regions for three years and rising only in the west, only after the invasion. The gap is the AIS side falling, not the radar side rising. At western ports the post-break radar deviation is $+0.07$ log points while the AIS record falls by $0.66$, and in the east the two move together. Two mechanical checks leave it in place. The pair-sum variable does rise with the interval between radar scenes, about five percent per additional day within port, but adding the interval as a covariate moves the gap to $+0.84$ ($p = 0.046$), and replacing the radar side with the count-frozen predictions leaves it at $+0.79$ ($p = 0.060$). The largest port-level increases are 2.5 log points at the Kavkaz Oil Terminal, the transshipment anchorage in the Kerch Strait, and 1.5 at Vysotsk, a port that \citet{fernandez2025charting} identify by name.

The gap says the radar sees activity in the west that the AIS record loses. The timing and the geography point to berth work by vessels with transponders off, and we read the gap as consistent with the dark-shipping account: it establishes that the radar retains activity the AIS record loses, and we do not put a volume on it. The implication for the design is directional: a version of \Cref{sec:russia} built on AIS alone could overstate the eastward shift by missing western activity, an exposure the radar does not carry.

\subsection{Both directions and the December 2022 embargo}

\begin{figure}[htbp!]
  \centering
  \begin{subfigure}[t]{0.32\textwidth}
    \centering
    \includegraphics[width=\textwidth]{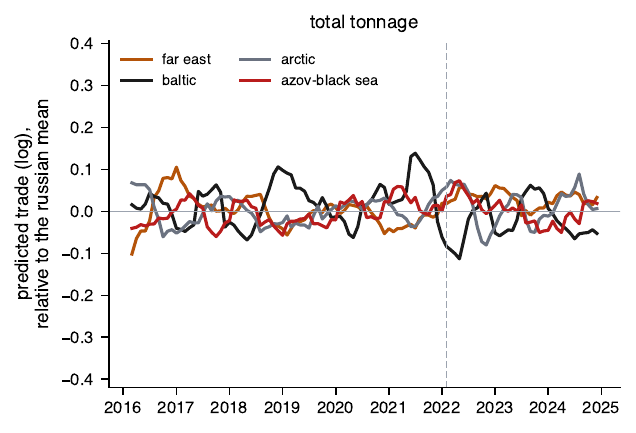}
    \caption{Total tonnage}\label{fig:rus_total}
  \end{subfigure}
  \hfill
  \begin{subfigure}[t]{0.32\textwidth}
    \centering
    \includegraphics[width=\textwidth]{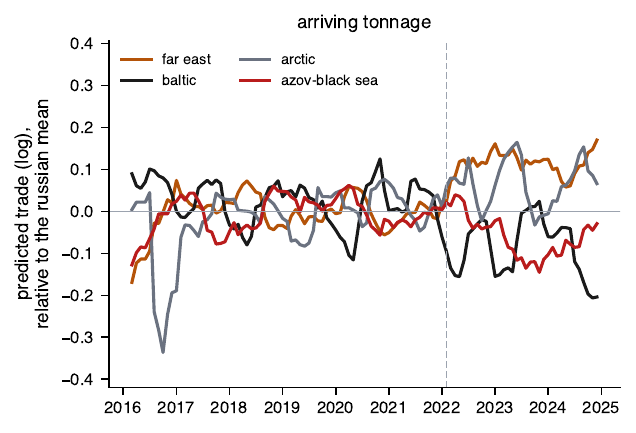}
    \caption{Arriving tonnage}\label{fig:rus_inward}
  \end{subfigure}
  \hfill
  \begin{subfigure}[t]{0.32\textwidth}
    \centering
    \includegraphics[width=\textwidth]{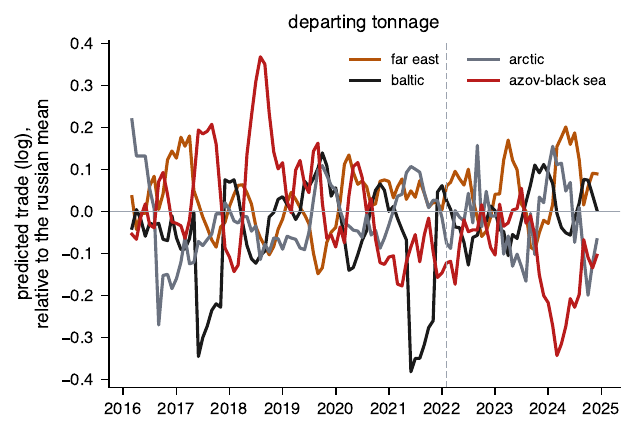}
    \caption{Departing tonnage}\label{fig:rus_outward}
  \end{subfigure}
  \caption{Russian ports by basin and direction of trade}
  \label{fig:russia_direction}
  \vspace{0.5em}
  \begin{minipage}{1\textwidth}
  \setstretch{1.15}
    \footnotesize
Notes: Each panel plots predicted log trade by basin, measured against each port's own
pre-sanction mean and then against the movement common to all Russian ports, six-month
moving average. All three panels use models trained on the U.S. panel and
share a vertical scale. The dashed line marks February 2022. Arriving tonnage in panel
(b) separates the basins most sharply, with the Far East rising and the Baltic falling,
which is the reallocation of imports from European to Asian suppliers. Departing
tonnage in panel (c) is the least well predicted quantity in \Cref{tab:direction} and
its panel should be read against that.
\end{minipage}
\end{figure}

Sanctions bound in both directions, on energy leaving Russia from December 2022 and on imports arriving from February 2022, so we run the design on each. \Cref{tab:russiadir} reports six models, one per direction and training panel, each tuned as \Cref{tab:direction} describes. The Far East coefficient is positive in all six. The total-tonnage estimate here, $+0.050$, differs from the $+0.206$ of \Cref{tab:russia} only in tuning: each model in this table is searched against its own direction's criterion, whereas the primary arm is searched against the paper's. We treat the primary arm as the headline and this table as the directional split, not as a second estimate of the total.

The clearest pattern is on arrivals. Trained on U.S. import weight, the estimate is $+0.183$ with a permutation $p = 0.103$, and the basins separate; they do not merely rise together. The Far East gains 0.109 relative to the common movement while the Baltic loses 0.097. That is the reallocation of arriving goods from European suppliers reached through the Baltic to Asian suppliers reached through the Pacific coast, and \Cref{fig:russia_direction} shows it.

Departures are weak, and \Cref{tab:direction} says why. Departing tonnage is the least well predicted quantity everywhere in the transfer matrix, so a null on departures is what this measure would produce whether or not Russian exports moved. We report it and do not read it as evidence.

\begin{table}[htbp]
\centering
\caption{The post-sanction split at the crude embargo}
\label{tab:embargo}
\begin{threeparttable}
\begin{tabular}{llcccc}
\toprule
& & \multicolumn{2}{c}{Far East against the rest} & & \\
\cmidrule(lr){3-4}
Target & Trained on & \makecell{Sanctions\\window} & \makecell{Embargo\\window}
& Difference & \makecell{Seasonal\\difference} \\
\midrule
Total & U.S. & \makecell{+0.067\\\footnotesize $p$ 0.326} & \makecell{+0.043\\\footnotesize $p$ 0.691} & \makecell{-0.024\\\footnotesize $p$ 0.730} & -0.033 \\
 & Europe & \makecell{+0.089\\\footnotesize $p$ 0.118} & \makecell{+0.073\\\footnotesize $p$ 0.205} & \makecell{-0.016\\\footnotesize $p$ 0.667} & -0.027 \\
\addlinespace
Arrivals & U.S. & \makecell{+0.186\\\footnotesize $p$ 0.065} & \makecell{+0.182\\\footnotesize $p$ 0.164} & \makecell{-0.005\\\footnotesize $p$ 0.964} & -0.009 \\
 & Europe & \makecell{+0.016\\\footnotesize $p$ 0.860} & \makecell{+0.051\\\footnotesize $p$ 0.392} & \makecell{+0.034\\\footnotesize $p$ 0.233} & +0.005 \\
\addlinespace
Departures & U.S. & \makecell{-0.020\\\footnotesize $p$ 0.842} & \makecell{+0.115\\\footnotesize $p$ 0.427} & \makecell{+0.136\\\footnotesize $p$ 0.179} & +0.078 \\
 & Europe & \makecell{+0.014\\\footnotesize $p$ 0.632} & \makecell{+0.052\\\footnotesize $p$ 0.214} & \makecell{+0.037\\\footnotesize $p$ 0.320} & +0.016 \\
\bottomrule
\end{tabular}
\begin{tablenotes}[flushleft]\footnotesize
\item Notes: The post-sanction indicator of \Cref{tab:russiadir} is replaced by two: the
sanctions window runs February to November 2022 and the embargo window from
December 2022, when the European Union ban
on seaborne Russian crude and the G7 price cap took effect. Each is measured against
the period before February 2022, so the difference is what the embargo added. Port and
month fixed effects throughout. No model is refitted; these are the same predicted
series as \Cref{tab:russiadir}. Permutation $p$ values reshuffle the basin
labels across the 35 ports, 3,000 draws, with both interactions recomputed under each
draw. The last column adds a basin-specific calendar-month effect, because the first
window contains no December or January and the second contains three of each. The
twelve window coefficients under the seasonal specification are in the replication
package.
\end{tablenotes}
\end{threeparttable}
\end{table}

Timing gives a second test. February 2022 is when the sanctions regime began, and it is not when the crude stopped going west. The European Union ban on seaborne Russian crude and the G7 price cap both took effect on 5 December 2022, and the ban on refined products on 5 February 2023. If the reallocation we report is crude finding new buyers, the gap should widen after the ban bound. \Cref{tab:embargo} replaces the post indicator with two and asks whether it did.

It does not widen. All six second-window coefficients are positive, three of the six exceed their first-window counterpart, and no difference approaches significance. The reading is that the reallocation was already in place when the embargo took effect, and the trade record says the same independently, since Russian crude began moving to Indian and Chinese buyers through the spring and summer of 2022 \citep{babina2023assessing, kilian2026impact}. December 2022 confirmed in law a redirection that had already happened. The first window is ten months and 189 observations, so the power to separate the two was limited, and we report the comparison as consistent with one break, not as evidence against a second.

One caution belongs with this. The first window contains no December and no January while the second contains three of each, and the Far East has a large arrival and departure premium in those two months in every year of the sample, so month fixed effects common to all Russian ports do not remove it. The last column of \Cref{tab:embargo} adds a basin-specific calendar-month effect. Nothing that matters moves, and every one of the twelve window coefficients is positive under it, whereas one was negative without it. \Cref{app:results} reports the seasonal placebo this rests on.

\subsection{Power}

\begin{figure}[htbp!]
  \centering
  \includegraphics[width=0.72\textwidth]{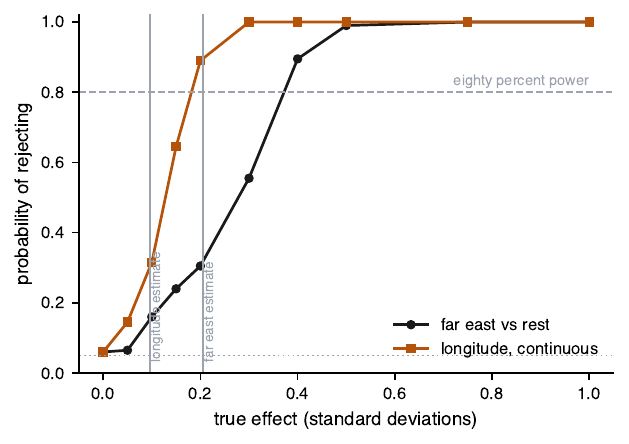}
  \caption{Rejection probability of the port-permutation test}
  \label{fig:power}
  \vspace{0.5em}
  \begin{minipage}{1\textwidth}
  \setstretch{1.15}
    \footnotesize
Notes: Rejection probability as a function of the true effect, for the two
designs of \Cref{sec:russia}. Two hundred simulated panels per point. Vertical
lines mark the estimates actually obtained.
\end{minipage}
\end{figure}

A non-significant estimate is uninformative unless the reader knows what the design could have found. At a true effect of zero the test in \Cref{tab:power} rejects at 6.0 percent in both designs, so the procedure is calibrated. Eighty percent power arrives at 0.40 standard deviations for the binary split and 0.20 for the continuous exposure, against estimates of 0.206 and 0.097, and \Cref{fig:power} plots both curves. At the effect we actually estimate this design rejects roughly one time in three. A permutation $p = 0.10$ is what an underpowered test returns when a real effect of that size is present, and it is not evidence of absence. We therefore report the result as directional, and we prefer the continuous specification because it delivers about twice the power of the binary one at the same effect size.

\subsection{Corroboration \label{sub:corroboration}}

Four things corroborate the direction independently of the estimate above. The first uses a reconstruction of the Association of Sea Ports of Russia series, which reports quay throughput including domestic and transit cargo and serves as a consistency check without being ground truth. Among the fourteen named ports with at least six months on each side of the break, all four Far Eastern ports rose while seven of ten western ports fell, giving a one-sided Fisher exact $p = 0.035$.

The second needs no model and no radar. PortWatch port calls cover the 25 matched Russian ports whether or not the radar does, so this check includes Vladivostok and the other ports the satellite panel loses after 2022. On estimated tonnage, seven of nine eastern ports rise against six of sixteen western ports, a one-sided Fisher exact $p = 0.063$; on port calls, six of nine against five of sixteen, $p = 0.098$. One caveat cuts against weighting it heavily: \Cref{sub:dark} shows the AIS record loses western activity after the break, which pushes this comparison toward the eastward pattern, so we report it beside the ASOP comparison, not above it.

The third bears on the design itself. Europe lets us run the operation \Cref{sec:russia} performs, ranking which ports moved most across a break, in a place where the realized change is observed. Training to the second quarter of 2022 and predicting the ten quarters that follow, the predicted change ranks the realized change across 105 European ports at $0.293$, with a 95 percent port-bootstrap interval of $[0.098, 0.471]$. The interval excludes zero, so ranking changes is not a vacuous operation for this model, which is what \Cref{sec:russia} requires of it.

The fourth is what partners report. Mirror statistics are the workhorse record of the sanctions literature \citep{crozet2020friendly, flach2024quantifying}, and they cannot measure port-level reallocation: no mirror record names a Russian port, only a third of post-break mirror value comes from a partner that files a mode of transport, and China, Korea and Japan file none, so the mode-resolved mirror is close to blind on the side this section is about. Where the mirror does speak it agrees, with trade rising 0.434 log points at Pacific partners and falling 0.450 at western ones, a gap of 0.884; \Cref{tab:mirror} in \Cref{app:results} reports the accounting.

One caveat attaches to the whole section. \Cref{tab:applicability} places 33 of the 35 Russian ports in the far quarter of the range the model was fitted on, though none beyond its maximum. This is the most extrapolated application in the paper, and the dissimilarity index is reported so that a reader can see it stated.

\section{Conclusion \label{sec:conclusion}}

Satellite imagery and port characteristics measure a port's trade without that port ever having reported, and the mapping carries in both directions between two continents whose trade statistics measure different things.  Applied to Russia the measure gives reallocation without contraction, with the Far East the only basin rising faster than the common movement. That direction is consistent with the reorientation of Russian trade toward Asia, and we obtain it at a design power we report and do not assume.

Three limitations bound the claim. Per-port levels are imprecise, typically by a factor of two or more, and the dissimilarity index says which ports lie at which end. Resolution is coarse, since the measure distinguishes ports differing by an order of magnitude and not ports differing by half. \Cref{sub:learning} says where the binding constraint is not. Performance stops improving in the number of training ports well before the panel runs out, and the target those ports supply is precise enough that a perfect model would score close to one. What is left is the description of the image. The most promising route is therefore a richer representation such as the random-feature encoding of \citet{rolf2021}, which replaces hand-built summary statistics with a general description of what a port looks like, and we would expect more from that than from another hundred labeled ports.

\clearpage
\bibliographystyle{ecta} \bibliography{ref}

\clearpage
\appendix

\section{Data Construction \label{app:data}}

\renewcommand{\thesection}{\Alph{section}}
\setcounter{table}{0}
\setcounter{figure}{0}
\renewcommand{\thetable}{A\arabic{table}}
\renewcommand{\thefigure}{A\arabic{figure}}

\subsection{The radar transform}

The Sentinel-1 collection distributes the VV band in decibels and the extraction applies a further logarithm before differencing, so only pixels brighter than 0 dB survive. Across 44 sampled port-months at eight ports spanning the range of window composition and size, the median survival is 5.4 percent, running from 0.9 percent at Erie to 15.7 percent at Los Angeles. The survivors are strong scatterers. Over permanent water, classified by the global surface water product of \citet{pekel2016} at above 50 percent occurrence, a median 73.6 percent of them belong to connected components of at least 20 pixels, which is ship-sized at 10 metres, against 4.3 percent in components of three pixels or fewer. A median 86.6 percent of surviving pixels sit on land, which includes the quay itself and everything berthed against it; only the remainder falls over water. The variable therefore acts as a bright-target change detector over the whole terminal.

\subsection{WPI fields that carry no information \label{app:const}}

Thirty of the ninety WPI fields are constant across the 68 U.S. ports and so contribute nothing to any specification fitted on that panel. They include every flag that names a cargo type. Eleven of the WPI's fourteen facility flags read ``Unknown'' at all 68 ports, among them \texttt{fac\_container}, \texttt{fac\_liquid\_bulk} and \texttt{fac\_oil\_terminal}. The three that vary, \texttt{fac\_wharves}, \texttt{fac\_anchorage} and \texttt{fac\_med\_mooring}, describe berth types and say nothing about cargo. This is why \Cref{tab:cargo} cuts on recorded terminal depths and handling equipment instead of cargo shares, which the WPI does not supply for these ports, and why \Cref{sub:cargo} has to go to Eurostat for composition.

Constant does not mean constant everywhere. The transfer matrix of \Cref{tab:transfer} uses the same ninety fields on both continents, but only six are constant across the 113 European ports and only five are constant in both panels. A model fitted on Europe therefore reads information from fields that are dead on the U.S. panel, and the reverse does not happen. That asymmetry is a property of the WPI's coverage, not of the ports, and it is one reason not to read the two cross-continent entries of \Cref{tab:transfer} as a matched pair.

The WPI also contains errors. Ust-Luga, which handled 130 million tonnes in the year before the sanctions, is recorded as ``Very Small''. We use characteristics as supplied and do not correct them, so any such error enters the results.

\subsection{The Russian panel}

The panel contains 35 ports, drawn from the 67 Russian entries of the WPI. The original construction kept the 28 recorded as Large, Medium, or Small. Seven more were added when we reran the satellite extraction with wider geographic coverage, namely Bukhta Vanino, Kavkaz Oil Terminal, Lomonosov, Primorsk, Shakhtersk, Ust-Luga, and Vostochnyy. All seven are recorded as ``Very Small'' and were excluded by the original size filter, which is the WPI error of \Cref{app:const} operating as a selection rule. Ports enter if they carry a WPI record and a usable radar observation, and we selected no port on any outcome. Restricting the primary arm's difference in differences to the original 28 ports gives $+0.159$ with a permutation $p = 0.167$, the same direction on a smaller panel with less power, so the estimate does not originate in the added terminals.

The Caspian ports are absent because the WPI does not list them. Astrakhan and Makhachkala have no WPI entry, so no Russian Caspian port could enter by this construction, and the basins of \Cref{tab:russia} therefore cover the Baltic, the Azov and Black Sea, the Arctic, and the Far East, which together account for all 35.

\subsection{Entity definitions in the ASOP comparison}

The reconstructed Association of Sea Ports of Russia series and the WPI do not name the same objects. Under Russian government order 1225-r, Saint Petersburg is a composite entity that includes Kronshtadt, Bronka and Lomonosov, which the WPI lists separately, and Vanino absorbs Sovetskaya Gavan. Our comparison in \Cref{sec:russia} counts ASOP entities as ASOP reports them, so Saint Petersburg is one observation there and three in the WPI. The count of four Far Eastern ports rising against seven of ten western ports falling is on the ASOP definition.

\subsection{Sensitivity to the window radius}

\begin{table}[htbp]
\centering
\caption{Sensitivity of the satellite variables to the window radius}
\label{tab:aoi}
\begin{threeparttable}
\begin{tabular}{lcccc}
\toprule
 & \multicolumn{2}{c}{Rank correlation with 3 km} & \multicolumn{2}{c}{Median ratio to 3 km} \\
\cmidrule(lr){2-3}\cmidrule(lr){4-5}
Variable & 2 km & 4 km & 2 km & 4 km \\
\midrule
SAR pair difference & 0.925 & 0.741 & 0.55 & 1.34 \\
VH backscatter & 0.961 & 0.976 & 1.00 & 0.99 \\
Lights, mean & 0.935 & 0.985 & 1.17 & 0.84 \\
Lit area share & 0.895 & 0.938 & 1.00 & 1.00 \\
\bottomrule
\end{tabular}
\begin{tablenotes}[flushleft]\footnotesize
\item Notes: Twenty U.S. ports spanning the size range, six months, recomputed at
each radius on the same collections and reducers as the production extraction. The
ordering of ports is preserved, with rank correlations of 0.74 to 0.99. The weakest
entry is the SAR pair difference at 4 km, which is a sum over the window and therefore
grows with area, as the ratio column shows. Since the paper claims a ranking and not
a level, sensitivity of the level to the window is not a threat to the result.
\end{tablenotes}
\end{threeparttable}
\end{table}

\Cref{tab:aoi} recomputes the satellite variables at 2 and 4 kilometers for twenty ports spanning the size range. The ordering of ports survives, with rank correlations against the 3 kilometer window of 0.74 to 0.99. Levels move, most for the SAR pair difference, which is a sum over the window and grows with its area. The paper claims a ranking, so the level sensitivity does not bear on the result.

\subsection{Panel coverage}

\Cref{tab:coverage} reports how much of each panel carries a usable radar observation, and how that changed when Sentinel-1B failed in December 2021. Coverage rises for the U.S. and European panels because the 2016 ramp-up of the constellation costs more observations than the 2021 loss of one satellite, and falls only for Russia.\footnote{A second constellation change falls inside the sample. Sentinel-1C launched on 5 December 2024, the last of the 108 months. It did not change the data we use. Mean usable scenes per port-month in that month are 5.27 for the U.S. panel against 4.78 over 2022-01 to 2024-11, 1.3 standard deviations of the ordinary monthly spread in the two-satellite era, and the European and Russian panels move by 0.4 and $-0.1$ standard deviations. One month of 108 either way.}

\begin{table}[htbp]
\centering
\caption{Panel coverage and the Sentinel-1B failure}
\label{tab:coverage}
\begin{threeparttable}
\begin{tabular}{lccccc}
\toprule
Panel & Ports & \makecell{Port-\\periods} & Span & \makecell{Radar\\coverage} & \makecell{After\\Dec.\ 2021} \\
\midrule
U.S. customs & 68 & 7,343 & 2016--2024 & 0.892 & 0.959 \\
Europe quay & 215 & 23,220 & 2016--2024 & 0.961 & 0.967 \\
Russia & 35 & 3,780 & 2016--2024 & 0.867 & 0.675 \\
\bottomrule
\end{tabular}
\begin{tablenotes}[flushleft]\footnotesize
\item Notes: Radar coverage is the share of port-periods with a usable Sentinel-1
observation. Of the 215 European ports with a radar extraction, 113 also carry a
usable Eurostat target, which is the sample the transfer results use.
Sentinel-1B failed in December 2021, halving the revisit rate.
Coverage nonetheless \emph{rises} for the U.S.\ and European panels, because the
2016 ramp-up of the constellation costs more observations than the 2021 loss. Only
the Russian panel falls, and it falls two months before the event studied in
\Cref{sec:russia}, which is a limit on the power of that section.
\end{tablenotes}
\end{threeparttable}
\end{table}

\section{Additional Results \label{app:results}}

\renewcommand{\thesection}{\Alph{section}}
\setcounter{table}{0}
\setcounter{figure}{0}
\renewcommand{\thetable}{B\arabic{table}}
\renewcommand{\thefigure}{B\arabic{figure}}

\subsection{Aggregate fit}

\begin{figure}[htbp]
  \centering
  \begin{subfigure}[t]{\textwidth}
    \centering
    \includegraphics[width=\textwidth]{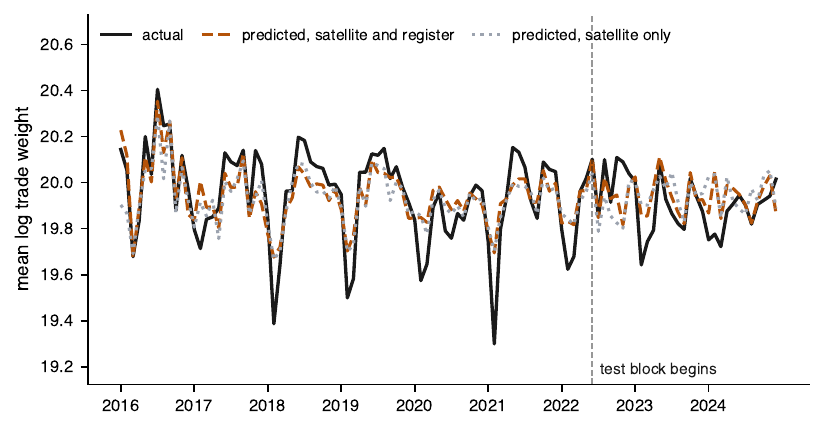}
    \caption{Mean log trade weight, actual and predicted}
    \label{fig:agg_trade_wgt}
  \end{subfigure}

  \vspace{0.6em}
  \begin{subfigure}[t]{\textwidth}
    \centering
    \includegraphics[width=\textwidth]{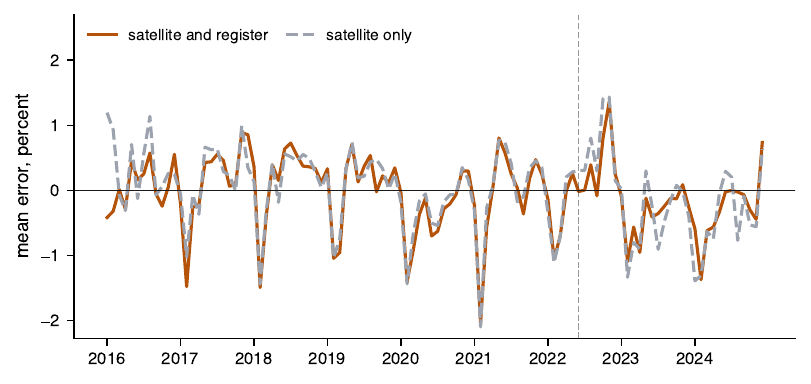}
    \caption{Error in the mean}
    \label{fig:agg_trade_resid_wgt}
  \end{subfigure}
  \caption{The panel average over time}
  \label{fig:agg_trade_prediction_wgt}
  \vspace{0.5em}
  \begin{minipage}{1\textwidth}
  \setstretch{1.15}
    \footnotesize
Notes: The mean of log trade weight over the ports carrying an observation in that
month, actual against the port-month predictions of the two specifications, averaged
for display only. The mean and not the sum. Ports carrying an observation rise from 13
to 66 over the sample, so a sum of log weight has an $R^2$ of 0.998 against the port
count alone and a figure of it would show that the model knows how many ports reported.
The vertical line marks June 2022, where the test block begins. Panel (b) is
$[(\text{actual}_t - \text{predicted}_t)/\text{actual}_t] \times 100$ computed on those
means, so one percent is about 0.2 log points.
\end{minipage}
\end{figure}

A pooled goodness-of-fit statistic computed at port-month level reaches 0.88 to 0.96. Almost all of that is variation between ports and almost none of it is movement inside any one of them, which is why the paper reports held-out ports instead of a pooled figure. \Cref{fig:agg_trade_prediction_wgt} shows what is left when the between-port part is taken out by averaging. The predicted mean follows the actual mean loosely and misses most of the deep January troughs, though it stays within about two percent of that mean, which on a log mean near twenty is around 0.4 log points. The next subsection quantifies what remains once between-port variation is removed.

\subsection{The time dimension at annual frequency}

\begin{table}[htbp]
\centering
\caption{The time dimension at annual frequency}
\label{tab:annualchange}
\begin{threeparttable}
\begin{tabular}{llcc}
\toprule
Panel & Variable & Port-years & $R^2$ of annual changes \\
\midrule
U.S. customs weight & SAR difference & 539 & 0.0006 \\
U.S. customs weight & VH backscatter & 442 & 0.0003 \\
U.S. customs weight & Lights, mean & 544 & 0.0128 \\
U.S. customs weight & Lights, s.d. & 544 & 0.0069 \\
U.S. customs weight & Lit area share & 544 & 0.0325 \\
U.S. customs value & SAR difference & 539 & 0.0003 \\
U.S. customs value & VH backscatter & 442 & 0.0000 \\
U.S. customs value & Lights, mean & 544 & 0.0034 \\
U.S. customs value & Lights, s.d. & 544 & 0.0018 \\
U.S. customs value & Lit area share & 544 & 0.0162 \\
Europe quay tonnage & SAR difference & 841 & 0.0001 \\
Europe quay tonnage & VH backscatter & 572 & 0.0005 \\
Europe quay tonnage & Lights, mean & 854 & 0.0001 \\
Europe quay tonnage & Lights, s.d. & 854 & 0.0000 \\
Europe quay tonnage & Lit area share & 854 & 0.0001 \\
\bottomrule
\end{tabular}
\begin{tablenotes}[flushleft]\footnotesize
\item Notes: Annual log changes within a port, with year fixed effects, so the column asks
whether a satellite change says anything beyond the year every port shared.
\citet{gibsonboe2021} report a within-county $R^2$ of 0.029 for annual U.S.\ county
GDP changes with the same sensor, falling to 0.005 where agriculture matters most.
The model's own predicted change reaches 0.0004, because it is fitted to
cross-sectional scale and does not use this component.
\end{tablenotes}
\end{threeparttable}
\end{table}

Restricted to movement inside a port, and measured at the annual frequency the nighttime-lights literature reports, the largest figure in \Cref{tab:annualchange} is 3.3 percent of the variation in annual trade changes explained by the lit-area share. \citet{gibsonboe2021} report a within-county $R^2$ of 0.029 for annual U.S. county GDP changes with the same sensor, falling to 0.005 in counties where agriculture matters most, so the figure here sits at the same level at a spatial unit finer than a county. The model's own predicted change reaches 0.0004, because it is fitted to cross-sectional scale and does not use that component. Six further specifications aimed at recovering within-port movement, covering differenced targets, de-seasonalized features, per-port time-series models, aggregation to basins and countries, lagged designs and vessel counts, did not change this. Our results should be read as cross-sectional.

\subsection{Ordering by port type}

\Cref{tab:cargo} cuts the held-out ports by the WPI fields that carry information about what a port handles. Ordering is best at ports with recorded bunkering and deep cargo piers, and it holds at ports with a recorded oil terminal, which is the group the Russian application depends on. It is weakest among the nine ports with no recorded oil terminal, where the interval is too wide to read. Those groups are named after WPI fields and not cargo shares, because the WPI supplies no shares, as \Cref{app:const} explains. Where composition can be measured, in Europe, it is, and \Cref{sub:cargo} reports that cut in the body.

\begin{table}[htbp]
\centering
\caption{Ordering accuracy by port type}
\label{tab:cargo}
\begin{threeparttable}
\begin{tabular}{lccc}
\toprule
Group & Ports & Rank correlation & \makecell{Pairs ordered\\correctly} \\
\midrule
Oil terminal depth recorded & 57 & \makecell{0.659 \\ \footnotesize [+0.464, +0.789]} & 73.9\% \\
No oil terminal depth recorded & 9 & \makecell{0.300 \\ \footnotesize [-0.487, +0.848]} & 61.1\% \\
Cargo pier depth above the median & 44 & \makecell{0.696 \\ \footnotesize [+0.491, +0.831]} & 75.6\% \\
Cargo pier depth below the median & 22 & \makecell{0.415 \\ \footnotesize [-0.005, +0.753]} & 64.9\% \\
Floating cranes recorded & 30 & \makecell{0.513 \\ \footnotesize [+0.172, +0.747]} & 68.5\% \\
Fuel oil bunkering recorded & 61 & \makecell{0.720 \\ \footnotesize [+0.577, +0.817]} & 76.3\% \\
\midrule
All ports & 66 & \makecell{0.668 \\ \footnotesize [+0.510, +0.786]} & 74.0\% \\
\bottomrule
\end{tabular}
\begin{tablenotes}[flushleft]\footnotesize
\item Notes: Held-out U.S. ports, weight target, from the predictions of
\Cref{tab:contribution}. Intervals are 4,000 bootstrap draws over the ports of the
group. The last column uses only pairs in which both ports belong to the group. The
WPI records no cargo shares and every one of its cargo facility flags is
``Unknown'' for every port, so each group is named after the field it was actually
built from. \Cref{app:results} reads the table.
\end{tablenotes}
\end{threeparttable}
\end{table}

\subsection{December and January in the Russian panel}

Splitting the post-sanction period more finely than \Cref{tab:embargo} is tempting, and it does not survive. Separating the refined products ban of February 2023 leaves a crude window of exactly two months, December 2022 and January 2023, and in that window the coefficients are large, up to $+0.506$ on departures with a permutation $p = 0.018$. They are seasonal. Treating the same two calendar months as the event in each of the eight years of the sample gives departure coefficients of $+0.09$ to $+0.38$ in the seven years when nothing happened. Pooling all years, the Far East sits 0.370 log points above its own mean in December and January on departures, against 0.162 for the Arctic, 0.066 for the Baltic and 0.027 for the Azov and Black Sea. A month effect common to all Russian ports removes none of that, because it falls differently across basins.

What survives the placebo is weaker. December 2022 is the largest or second-largest of the eight windows in all six specifications, and that holds when the ten Arctic ports are dropped. With eight windows a rank of first is a one-sided $p = 0.125$ in any one specification, and the six are the same predicted series cut six ways, not six independent tests. We record it and claim nothing from it.

\subsection{Ordering at a single date}

\begin{figure}[htbp]
  \centering
  \includegraphics[width=0.86\textwidth]{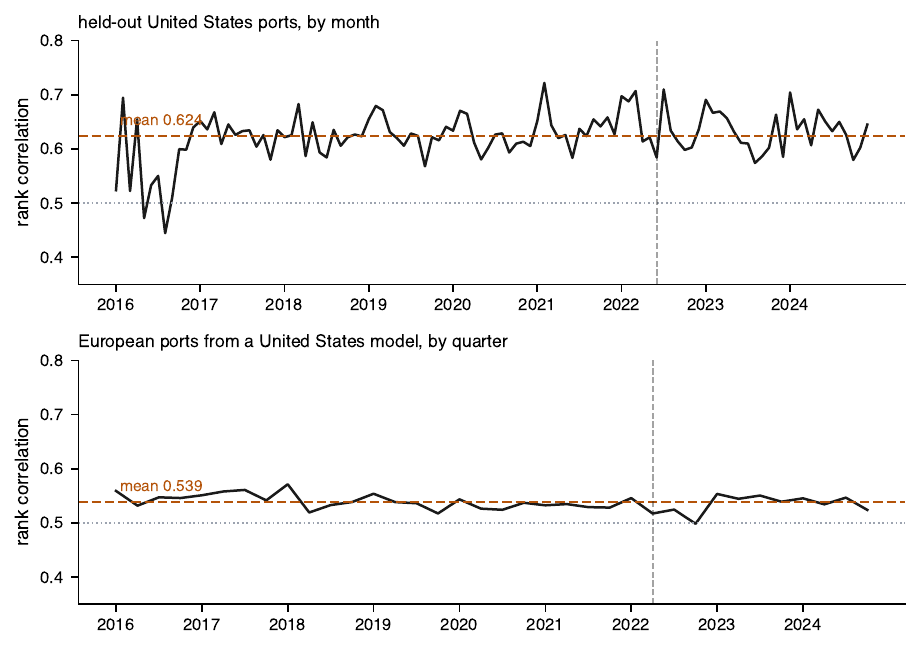}
  \caption{Port ordering at each date}
  \label{fig:perdate}
  \vspace{0.5em}
  \begin{minipage}{1\textwidth}
  \setstretch{1.15}
    \footnotesize
Notes: The rank correlation between actual and predicted trade computed separately at
each date, not pooled. The upper panel holds out one U.S. port at a time, so
no port contributes to its own prediction, and the lower panel orders the European
panel from a model trained on U.S. ports alone, which has seen none of them.
Dashed horizontal lines are the means across dates, the dotted line is 0.5, and the
vertical line marks the start of the test block. Dates with fewer than eight observed
ports are omitted. This is the operation \Cref{sec:russia} performs at each date, shown
where the answer is observed.
\end{minipage}
\end{figure}

\Cref{sec:russia} ranks the Russian ports at each date and asks whether that ranking shifted, so what matters there is not the pooled ordering but the ordering within one date. \Cref{fig:perdate} shows that operation where the answer is observed. Held-out U.S. ports are ordered at 0.624 on average across 108 months, below 0.5 in two of them and at least 0.44 in all of them. Ordering the European panel out of domain, one quarter at a time, gives 0.539 across 36 quarters and falls below 0.5 once.

Neither series decays where it would matter. The monthly figure averages 0.617 before the Sentinel-1B failure of December 2021 and 0.636 after it, and 0.620 in the training block against 0.632 in the test block. The quarterly transfer figure is flat at 0.54 throughout. The early months of 2016 are the weakest and are also the months with the fewest observed ports, which is the constellation still filling in, not the mapping failing.

\subsection{Splitting the sample period}

\begin{table}[htbp]
\centering
\caption{The ordering in each half of the sample period}
\label{tab:periodsplit}
\begin{threeparttable}
\begin{tabular}{lccc}
\toprule
Period & Port-months & Ports & Rank correlation \\
\midrule
2016--2019 & 2,681 & 68 & \makecell{0.681 \\ \footnotesize [+0.508, +0.810]} \\[2pt]
2020--2024 & 3,884 & 68 & \makecell{0.723 \\ \footnotesize [+0.577, +0.829]} \\[2pt]
\bottomrule
\end{tabular}
\begin{tablenotes}[flushleft]\footnotesize
\item Notes: Each half receives its own 200-trial search under the protocol of
\Cref{tab:transfer} and its own leave-one-port-out pass, 1,068 model fits each. The
two halves rank the same 68 ports at 0.961. The full-period figure
in \Cref{tab:transfer} is 0.710, so the ordering does not depend on the sample
period.
\end{tablenotes}
\end{threeparttable}
\end{table}

\Cref{tab:periodsplit} repeats the leave-one-port-out exercise separately in each half of the sample, tuning each half on its own. The two halves rank the same 68 ports at 0.961, so the ordering does not depend on the sample period.

\subsection{Region holdout \label{sub:region}}

\begin{table}[htbp]
\centering
\caption{Region holdout}
\label{tab:region}
\begin{threeparttable}
\begin{tabular}{lcccc}
\toprule
Region held out & Ports & Rank correlation & Mean $|$error$|$ & Level offset \\
\midrule
East Coast & 21 & \makecell{0.719\\\footnotesize [+0.317, +0.912]} & 0.672 & +0.169 \\
Gulf Coast & 18 & \makecell{0.622\\\footnotesize [+0.228, +0.859]} & 2.102 & +1.000 \\
West Coast & 17 & \makecell{0.664\\\footnotesize [+0.237, +0.885]} & 1.101 & +0.263 \\
Lake Erie & 5 & \makecell{0.100\\\footnotesize [-1.000, +1.000]} & 1.671 & -1.156 \\
Alaska & 2 & \footnotesize too few ports & 0.266 & +0.266 \\
Hawaii & 2 & \footnotesize too few ports & 1.378 & -1.378 \\
Lake Michigan & 2 & \footnotesize too few ports & 1.152 & -1.152 \\
Lake Superior & 1 & \footnotesize too few ports & 0.036 & +0.036 \\
\bottomrule
\end{tabular}
\begin{tablenotes}[flushleft]\footnotesize
\item Notes: Every port of a region held out together, weight target, the configuration of \Cref{tab:transfer}
read from file and held fixed across folds so that a held-out region
cannot influence its own hyperparameters. Intervals are 4,000 bootstrap draws over the
ports of the region. Holding out one port at a time over all 68 gives 0.710 for
comparison. A region with fewer than five ports gets no rank correlation; with only two
ports it takes the values plus or minus one. That is why Alaska and Hawaii are
listed with their errors instead, and it is the reason a Hawaii exercise could not
establish transfer. Level offset is the mean of actual minus predicted over the
region's ports.
\end{tablenotes}
\end{threeparttable}
\end{table}

\Cref{tab:region} holds out every U.S. region in turn, so a model that has seen no port of a region is asked to rank all of them. It costs almost nothing. The East Coast, twenty-one ports withdrawn together, comes back at 0.719 against 0.710 when ports are withdrawn one at a time, and the West Coast and the Gulf at 0.664 and 0.622.

Four regions have fewer than five ports and get no rank correlation, because with only two ports a rank correlation is plus or minus one whatever the model does. Alaska is one of these, and what can be said is that its two ports carry a mean absolute error of 0.266 log points against a median of 1.126 across regions. It is the only high-latitude U.S. region and so the closest domestic analogue to the Russian Arctic, which is why we name it.

The level offsets in the last column run from $-1.378$ to $+1.000$ log points. A region withdrawn as a block carries a common offset that a single port does not, the same thing \Cref{tab:levelgap} finds across continents and the same reason the paper reports rank alone.

\subsection{Model class}

\Cref{tab:modelclass} gives each of nine model classes its own 300-trial search under one protocol and then scores it by leave-one-port-out. The leading class separates from only one of eight rivals on weight and two of eight on value, and ridge regression sits inside the leader's interval, so no conclusion in this paper turns on the learner. The gradient-boosting entry here reaches 0.670 on weight against the 0.710 diagonal of \Cref{tab:transfer}. The sample and the objective are the same, the two searches are independent 300-trial draws, and the difference sits well inside either bootstrap interval.

\begin{table}[htbp]
\centering
\caption{Nine model classes, each separately tuned}
\label{tab:modelclass}
\begin{threeparttable}
\setlength{\tabcolsep}{4pt}
\begin{tabular}{lcccc}
\toprule
 & \multicolumn{2}{c}{Weight} & \multicolumn{2}{c}{Value} \\
\cmidrule(lr){2-3}\cmidrule(lr){4-5}
Class & Rank corr. & 95\% interval & Rank corr. & 95\% interval \\
\midrule
Random forest & 0.691 & [0.520, 0.810] & 0.724 & [0.593, 0.815] \\
Histogram boosting & 0.672 & [0.508, 0.791] & 0.688 & [0.515, 0.806] \\
Gradient boosting (XGBoost) & 0.670 & [0.495, 0.790] & 0.715 & [0.575, 0.816] \\
Extremely randomized trees & 0.649 & [0.456, 0.785] & 0.729 & [0.615, 0.809] \\
Ridge & 0.626 & [0.465, 0.748] & 0.680 & [0.544, 0.770] \\
Support vector, RBF & 0.614 & [0.447, 0.737] & 0.699 & [0.571, 0.782] \\
Elastic net & 0.598 & [0.420, 0.729] & 0.673 & [0.529, 0.771] \\
Nearest neighbors & 0.521 & [0.338, 0.654] & 0.571 & [0.397, 0.696] \\
Neural network & 0.189 & [-0.058, 0.414] & 0.339 & [0.116, 0.533] \\
\bottomrule
\end{tabular}
\begin{tablenotes}[flushleft]\footnotesize
\item Notes: Each class receives its own 300-trial Optuna search against a five-fold
port-grouped objective, then leave-one-port-out over all U.S. ports; 25,187 model
fits in total. Intervals are 4,000 bootstrap draws over ports, the same resampled
ports entering every class in a draw. Paired against the leading class, the leader
is distinguishable from only one of eight rivals on weight and two of eight on
value. Nothing in this paper depends on the choice of learner.
\end{tablenotes}
\end{threeparttable}
\end{table}

\subsection{Estimating on a prediction}

\begin{table}[htbp]
\centering
\caption{Estimation on a predicted outcome}
\label{tab:montecarlo}
\begin{threeparttable}
\begin{tabular}{lccc}
\toprule
Error structure & \makecell{Largest \\ $|$bias$|$} &
\makecell{Bias at \\ $\gamma=0.20$} &
\makecell{Relative standard \\ deviation} \\
\midrule
As calibrated & 0.0017 & -0.0017 & 2.84 \\
Port component doubled & 0.0010 & +0.0002 & 2.86 \\
Period component doubled & 0.0017 & -0.0005 & 5.47 \\
Error shifts after the break, every port & 0.0018 & -0.0007 & 2.84 \\
Error shifts after the break, east only & 0.1023 & +0.0983 & 2.82 \\
\bottomrule
\end{tabular}
\begin{tablenotes}[flushleft]\footnotesize
\item Notes: Simulation. 4,000 panels per grid point, five
true effects from zero to 0.40 log points. The panel is the Russian one in shape,
35 ports of which
12 eastern and
108 months with the break at month
74. The prediction error has a port component
and a period component, set from the held-out U.S. exercise at
1.831 and
0.796 log points against an outcome
standard deviation of 2.248. The last column
is the standard deviation of the estimate divided by what an outcome without
prediction error would give.
\end{tablenotes}
\end{threeparttable}
\end{table}

Regressing on a predicted outcome is unusual enough to warrant a formal answer. \Cref{tab:montecarlo} gives one by simulation, with the two parts of the prediction error set from the held-out U.S. exercise. The estimator is unbiased when the error does not move with the treatment: across five true effects the largest bias is 0.0017 log points. The port component, the larger of the two, is absorbed by the port fixed effects however big it is, and an error that shifts for every port after the break is absorbed by the month fixed effects. What the prediction costs is precision, a standard deviation 2.8 times what an outcome without prediction error would give, which is the power \Cref{tab:power} reports.

\subsection{Mirror statistics}

\begin{table}[htbp]
\centering
\caption{Partner mirror statistics}
\label{tab:mirror}
\begin{threeparttable}
\begin{tabular}{lcc}
\toprule
\multicolumn{3}{l}{\emph{Panel A. What the mirror resolves}} \\
\midrule
Reporting partners & \multicolumn{2}{c}{147} \\
Partners filing a mode of transport & \multicolumn{2}{c}{78} \\
Post-break value that is mode resolved & \multicolumn{2}{c}{0.333} \\
Records naming a Russian port & \multicolumn{2}{c}{0} \\
\addlinespace
Pacific partners filing a sea mode & \multicolumn{2}{c}{5 of 12} \\
Western partners filing a sea mode & \multicolumn{2}{c}{15 of 25} \\
\midrule
\multicolumn{3}{l}{\emph{Panel B. Change after February 2022, by the coast that serves
the partner}} \\
\midrule
& All reporters & Balanced \\
Pacific & +0.434 & +0.452 \\
Western & -0.450 & -0.412 \\
Overland & -0.303 & +0.911 \\
Other & +0.227 & +0.300 \\
\addlinespace
Pacific minus western & +0.884 & +0.865 \\
\addlinespace
Pacific share of the total & 0.287 to 0.431 & 0.331 to 0.467 \\
Western share of the total & 0.415 to 0.258 & 0.468 to 0.278 \\
\bottomrule
\end{tabular}
\begin{tablenotes}[flushleft]\footnotesize
\item Notes: United Nations Comtrade, what partners report trading with Russia, 2017 to 2024,
aggregates excluded so members are not counted twice. A partner's total is its
all-modes row where it files one and the sum over its mode rows where it does not.
Panel B compares the mean month after February 2022 with the mean month before, in
log. Partners are grouped by the Russian coast that serves them, which is geography
and not a routing record. Balanced keeps the 66 partners filing in every month, which
removes reporters leaving as trade with Russia ends. \Cref{sub:corroboration} reads
the table. These totals are all modes, so the Pacific figure includes rail, road and
pipeline trade with China and does not measure seaborne trade through Far Eastern
ports.
\end{tablenotes}
\end{threeparttable}
\end{table}

\Cref{tab:mirror} measures what partner mirror statistics resolve. No mirror record names a Russian port, so the ordering of ports cannot be recovered from them at any level of accuracy. Nor, for most of the relevant trade, can the mode of transport: only a third of post-break mirror value comes from a partner that files one, and among the partners served by the Pacific coast just five of twelve do, with China, Korea and Japan filing none. The mode-resolved mirror sees the western side and is close to blind on the eastern side, the side \Cref{sec:russia} is about.

Where the mirror does speak it agrees. Grouping partners by the Russian coast that serves them, trade with Pacific partners rises by 0.434 log points from the mean pre-break month to the mean post-break month while trade with western partners falls by 0.450, a gap of 0.884, and the Pacific share of the mirrored total moves from 0.287 to 0.431 while the western share moves from 0.415 to 0.258. Restricting to the 66 partners that file in every month, which removes reporters leaving as trade ends, gives a gap of 0.865. These totals are all modes, so the Pacific figure carries rail, road and pipeline trade with China and is not a measure of seaborne trade through Far Eastern ports. It corroborates the direction and it is not a substitute for measuring it.

\subsection{Placebo break dates}

\begin{table}[htbp]
\centering
\caption{The change-ranking exercise at placebo break dates}
\label{tab:placebo}
\begin{threeparttable}
\begin{tabular}{lccc}
\toprule
Break date & Training quarters & Ports & Rank correlation of the change \\
\midrule
2018-Q1 & 8 & 109 & \makecell{0.122 \\ \footnotesize [-0.076, +0.322]} \\[2pt]
2018-Q3 & 10 & 110 & \makecell{0.241 \\ \footnotesize [+0.049, +0.417]} \\[2pt]
2019-Q1 & 12 & 108 & \makecell{0.200 \\ \footnotesize [-0.012, +0.395]} \\[2pt]
2019-Q3 & 14 & 110 & \makecell{0.169 \\ \footnotesize [-0.049, +0.367]} \\[2pt]
2020-Q1 & 16 & 111 & \makecell{0.224 \\ \footnotesize [+0.006, +0.419]} \\[2pt]
2020-Q3 & 18 & 111 & \makecell{0.214 \\ \footnotesize [+0.020, +0.394]} \\[2pt]
2021-Q1 & 20 & 109 & \makecell{0.224 \\ \footnotesize [+0.027, +0.418]} \\[2pt]
2021-Q3 & 22 & 109 & \makecell{0.173 \\ \footnotesize [-0.024, +0.358]} \\[2pt]
\textbf{2022-Q3} & 26 & 105 & \makecell{\textbf{0.293} \\ \footnotesize [+0.098, +0.471]} \\[2pt]
\bottomrule
\end{tabular}
\begin{tablenotes}[flushleft]\footnotesize
\item Notes: Each row trains on everything before the stated date and predicts the ten
quarters after it, so the number of ports and the length of the prediction problem do
not move with the date. Intervals are 4,000 bootstrap draws over ports. The bold row
is the sanctions break, the quarter in which the ten-quarter window opens. The 2020 rows are not clean
placebos, since the pandemic was a genuine continent-scale shock, and we report
them. \Cref{sec:russia} reads the table.
\end{tablenotes}
\end{threeparttable}
\end{table}

\Cref{tab:placebo} repeats the change-ranking exercise of \Cref{sub:corroboration} at eight earlier break dates, two of which fall in the pandemic and are reported as such, holding the prediction window at ten quarters throughout so that the length of the prediction problem does not move with the date. The placebo dates return a median of 0.207. We use this exercise to license the operation and not to date the break.

\begin{table}[htbp]
\centering
\caption{The estimating equation at placebo break dates}
\label{tab:placebo_est}
\begin{threeparttable}
\begin{tabular}{lcc}
\toprule
Placebo break & East $\times$ Post & Permutation $p$ \\
\midrule
February 2018 & $-0.029$ & 0.622 \\
February 2019 & $-0.009$ & 0.875 \\
February 2020\tnote{a} & $-0.003$ & 0.967 \\
February 2021\tnote{a} & $+0.017$ & 0.883 \\
\midrule
February 2022 (the break) & $+0.206$ & 0.100 \\
\bottomrule
\end{tabular}
\begin{tablenotes}[flushleft]\footnotesize
\item Notes: \Cref{eq:did} with the post indicator moved to each placebo date,
estimated on the sample ending 2022-01 so the real break never enters. Port and
month effects throughout; permutation $p$ over 3,000 reassignments of the basin
labels. A systematic prediction error drifting east at arbitrary dates would
appear here.
\item[a] The pandemic falls inside these windows, as in
\Cref{tab:placebo}.
\end{tablenotes}
\end{threeparttable}
\end{table}

\Cref{tab:placebo_est} answers a different question with the same device. It moves the post indicator of \Cref{eq:did} to four placebo dates on the pre-break sample, so it tests the estimating equation itself for an error that drifts east at arbitrary dates. The four estimates sit between $-0.029$ and $+0.017$ against $+0.206$ at the break.

\subsection{Tables behind the figures}

Each table in this subsection reports the estimates behind a figure or a number stated in the body.

\begin{table}[htbp]
\centering
\caption{The cross-continent level gap}
\label{tab:levelgap}
\begin{threeparttable}
\begin{tabular}{lccccc}
\toprule
 & Offset & \multicolumn{3}{c}{Levels $R^2$} & Rank \\
\cmidrule(lr){3-5}
Trained on, tested on & (log) & raw & after intercept & after slope & corr. \\
\midrule
U.S. to Europe & -12.44 & -79.56 & +0.320 & +0.350 & 0.581 \\
Europe to the U.S. & +12.25 & -31.52 & +0.300 & +0.316 & 0.624 \\
Europe, one port held out & +0.05 & +0.28 & +0.285 & +0.286 & 0.591 \\
U.S., one port held out & +0.07 & +0.41 & +0.410 & +0.410 & 0.710 \\
\bottomrule
\end{tabular}
\begin{tablenotes}[flushleft]\footnotesize
\item Notes: Computed on port means. ``After intercept'' shifts the prediction by the
difference in means and ``after slope'' also rescales it, which is the
two-parameter figure of \Cref{tab:transfer}. Off the diagonal a single additive
shift removes 97.8 and 99.2 percent of the raw squared error, and the remaining
rescaling removes less than one tenth of one percent. On the diagonal the offset is
0.07 log points and removes almost nothing. The rank correlation is unchanged by
either step because both are monotone. The cross-continent figure is therefore a
statement about two targets measured on different scales and not about the model
failing to order ports.
\end{tablenotes}
\end{threeparttable}
\end{table}

\begin{table}[htbp]
\centering
\caption{Transfer by direction of trade}
\label{tab:direction}
\begin{threeparttable}
\begin{tabular}{lccc}
\toprule
 & \multicolumn{3}{c}{Rank correlation of port means} \\
\cmidrule(lr){2-4}
Trained on, tested on & Total & Inward & Outward \\
\midrule
U.S., one port held out & \makecell{0.710 \\ \footnotesize [0.562, 0.817]} & \makecell{0.781 \\ \footnotesize [0.677, 0.850]} & \makecell{0.483 \\ \footnotesize [0.277, 0.654]} \\[2pt]
U.S. to Europe & \makecell{0.581 \\ \footnotesize [0.436, 0.686]} & \makecell{0.604 \\ \footnotesize [0.468, 0.717]} & \makecell{0.341 \\ \footnotesize [0.153, 0.509]} \\[2pt]
Europe to the U.S. & \makecell{0.624 \\ \footnotesize [0.426, 0.774]} & \makecell{0.661 \\ \footnotesize [0.498, 0.780]} & \makecell{0.559 \\ \footnotesize [0.350, 0.718]} \\[2pt]
Europe, one port held out & \makecell{0.591 \\ \footnotesize [0.453, 0.698]} & \makecell{0.635 \\ \footnotesize [0.514, 0.732]} & \makecell{0.496 \\ \footnotesize [0.328, 0.632]} \\[2pt]
\bottomrule
\end{tabular}
\begin{tablenotes}[flushleft]\footnotesize
\item Notes: Neither direction is estimated. U.S. customs data report import and
export vessel weight as separate series whose sum is the total, and the Eurostat
series carries a direction dimension with values Inwards and Outwards. Each column is
a target in its own right, each model is tuned on its own by 300 Optuna trials against
the RMSE of port means under five-fold grouping over ports, and intervals are 4,000
bootstrap draws over ports. Inward tonnage is better predicted than the total in every
combination and outward tonnage worse in every one. Cargo arriving is discharged and dwells
in the terminal, while cargo leaving arrives from inland and is loaded, so the two
spend very different amounts of time inside the window the satellite observes.
\end{tablenotes}
\end{threeparttable}
\end{table}

\begin{table}[htbp]
\centering
\caption{Ordering accuracy by size separation}
\label{tab:resolution}
\begin{threeparttable}
\begin{tabular}{lcc}
\toprule
Pairs of held-out ports & Pairs & Ordered correctly \\
\midrule
within a factor of 1.6 & 320 & 53.4\% \\
1.6 to 3-fold & 439 & 64.9\% \\
3 to 10-fold & 653 & 75.3\% \\
10 to 20-fold & 250 & 85.6\% \\
more than 20-fold & 483 & 88.2\% \\
\midrule
All pairs & 2,145 & 74.0\% \\
\midrule
\multicolumn{3}{l}{\textit{Both ports in the same size tercile}} \\
Small ports & 231 & 52.8\% \\
Medium ports & 231 & 60.2\% \\
Large ports & 231 & 61.0\% \\
\bottomrule
\end{tabular}
\begin{tablenotes}[flushleft]\footnotesize
\item Notes: All pairs of the 66 held-out U.S. ports, from the leave-one-port-out
predictions of \Cref{tab:contribution}, binned by how far apart the two ports actually
are. Fifty percent is chance. The measure separates ports that differ by an order of
magnitude and does not separate ports that differ by half.
\end{tablenotes}
\end{threeparttable}
\end{table}

\begin{table}[htbp]
\centering
\caption{The dissimilarity index}
\label{tab:applicability}
\begin{threeparttable}
\begin{tabular}{lc}
\toprule
 & Value \\
\midrule
\multicolumn{2}{l}{\textit{Panel A. Predicting held-out error, U.S. ports}} \\
Rank correlation, full panel & +0.295 \\
\quad $p$ & 0.015 \\
Mean over 60 random half-splits & +0.251 \\
Splits with a positive correlation & 56 of 60 \\
\midrule
\multicolumn{2}{l}{\textit{Panel B. Position of the Russian ports}} \\
Ports beyond the maximum dissimilarity seen in training & 0 of 35 \\
Ports in the most dissimilar quarter of the training range & 33 of 35 \\
\bottomrule
\end{tabular}
\begin{tablenotes}[flushleft]\footnotesize
\item Notes: The dissimilarity index of \citet{meyer2021} measures a port's distance in
predictor space from the training data, weighting predictors by their importance in
the fitted model, and is computed from the configuration of \Cref{tab:transfer}. The
top panel validates it out of sample: across 60 random half-splits of the U.S. panel it predicts held-out error, and \Cref{fig:di} plots the relation on the
full panel. The bottom panel places the Russian ports. None requires the model to
extrapolate beyond the range it was fitted on, but almost all of them sit in the far
end of that range, which is the caveat that attaches to \Cref{sec:russia}. We do not
report error multiples by quartile, because a quartile of seventeen ports is too few
for its median to be stable, and the relation is continuous, not stepped.
\end{tablenotes}
\end{threeparttable}
\end{table}

\begin{table}[htbp]
\centering
\caption{The Russian estimate by direction of trade}
\label{tab:russiadir}
\begin{threeparttable}
\begin{tabular}{llccrrrr}
\toprule
 & Trained & Far East & & \multicolumn{4}{c}{Relative change by basin} \\
\cmidrule(lr){5-8}
Direction & on & $\times$ Post & $p$ & Far East & Arctic & Azov-Black & Baltic \\
\midrule
Total & US & +0.050 & 0.598 & +0.022 & -0.006 & +0.004 & -0.020 \\
Total & Europe & +0.077 & 0.169 & +0.042 & -0.048 & +0.010 & -0.033 \\
Inward & US & +0.183 & 0.103 & +0.109 & +0.023 & -0.046 & -0.097 \\
Inward & Europe & +0.041 & 0.539 & +0.018 & +0.008 & +0.014 & -0.031 \\
Outward & US & +0.077 & 0.541 & +0.050 & +0.022 & -0.121 & +0.004 \\
Outward & Europe & +0.041 & 0.240 & +0.021 & +0.034 & +0.007 & -0.037 \\
\bottomrule
\end{tabular}
\begin{tablenotes}[flushleft]\footnotesize
\item Notes: Six models, one per direction and training panel, each tuned as
\Cref{tab:direction} describes and then applied to the 35 Russian ports. The Far East
coefficient is positive in all six.
Sanctions bound in both directions, on energy leaving from December 2022 and on
imports arriving from February 2022. The inward model trained on the U.S.
gives the clearest pattern, with the Far East rising and the Baltic falling relative
to the common movement, which is the reallocation of arriving goods from European to
Asian suppliers. The outward models are weak, and \Cref{tab:direction} shows why:
outward tonnage is the least well predicted quantity in every combination.
\end{tablenotes}
\end{threeparttable}
\end{table}

\begin{table}[htbp]
\centering
\caption{The radar--AIS gap at Russian ports after February 2022}
\label{tab:portwatch_gap}
\begin{threeparttable}
\begin{tabular}{lcccc}
\toprule
 & West $\times$ Post & \makecell{+ West $\times$ \\ calendar month} &
 \makecell{Sanctions \\ window} & \makecell{Embargo \\ window} \\
\midrule
Tonnage gap & \makecell{0.786 \\ \footnotesize(0.063)} & \makecell{0.793 \\ \footnotesize(0.065)} & \makecell{0.618 \\ \footnotesize(0.023)} & \makecell{0.856 \\ \footnotesize(0.080)} \\[4pt]
Port-call gap & \makecell{0.727 \\ \footnotesize(0.040)} & \makecell{0.733 \\ \footnotesize(0.038)} & \makecell{0.499 \\ \footnotesize(0.027)} & \makecell{0.822 \\ \footnotesize(0.039)} \\
\bottomrule
\end{tabular}
\begin{tablenotes}[flushleft]\footnotesize
\item Notes: The outcome is the gap between the radar index and the AIS record at the
same port and month: the tonnage gap is predicted log trade minus log PortWatch
tonnage, and the port-call gap subtracts log one plus port calls instead. Every specification carries port
and month effects; West is the Baltic and Azov--Black Sea basins, where
\citet{fernandez2025charting} locate dark-shipping activity, 13 of 23 ports and
1,165 port-months with radar and AIS both
observed, 2019-01 to 2024-12. The sanctions window is 2022-02 to 2022-11 and the
embargo window begins 2022-12. Permutation $p$-values over 3,000 reassignments of
the West label across ports in parentheses. A positive coefficient says the radar
sees activity in the west, relative to the east, that the AIS record loses after
the break.
\end{tablenotes}
\end{threeparttable}
\end{table}

\begin{table}[htbp]
\centering
\caption{Power of the port-permutation test}
\label{tab:power}
\begin{threeparttable}
\begin{tabular}{lcc}
\toprule
True effect (standard deviations) & Far East vs rest & longitude, continuous \\
\midrule
0.00 \textit{(size check)} & 0.060 & 0.060 \\
0.05  & 0.065 & 0.145 \\
0.10  & 0.160 & 0.315 \\
0.15  & 0.240 & 0.645 \\
0.20  & 0.305 & 0.890 \\
0.30  & 0.555 & 1.000 \\
0.40  & 0.895 & 1.000 \\
0.50  & 0.990 & 1.000 \\
0.75  & 1.000 & 1.000 \\
1.00  & 1.000 & 1.000 \\
\bottomrule
\end{tabular}
\begin{tablenotes}[flushleft]\footnotesize
\item Notes: Rejection rate of the port-permutation test in 200 simulated panels per
grid point. Each panel is built by two-way demeaning the real outcome to strip port and
period structure, drawing a fresh random treatment assignment of the same size as
the real one, and adding the stated effect. \Cref{sec:russia} reads the table.
\end{tablenotes}
\end{threeparttable}
\end{table}

\subsection{Vessel counts isolated over water}

Because the radar variable mixes water and land, we extracted a separate block counting bright pixels over permanent water only, grouped into ship-sized connected components, for all three panels. It does not improve prediction at an unseen port, and scrambling the block changes the score by 0.001; the replication package reports the arm scores.

\end{document}